\RequirePackage{filecontents}
\documentclass[a4paper, amsfonts, amssymb, amsmath, reprint, showkeys, nofootinbib, twoside]{revtex4-2}
\usepackage[english]{babel}
\usepackage[utf8]{inputenc}
\usepackage[colorinlistoftodos, color=green!40, prependcaption]{todonotes}
\usepackage[per-mode = fraction, separate-uncertainty = true]{siunitx}
\DeclareSIUnit{\lines}{lines}
\usepackage{amsthm}
\usepackage{mathtools}
\usepackage{physics}
\usepackage{xcolor}
\usepackage{graphicx}
\usepackage[left=23mm,right=13mm,top=35mm,columnsep=15pt]{geometry} 
\usepackage{adjustbox}
\usepackage{placeins}
\usepackage[T1]{fontenc}
\usepackage{lipsum}
\usepackage{csquotes}

\usepackage{siunitx}

\usepackage{xr}
\usepackage[pdftex, pdftitle={Article}, pdfauthor={Author}]{hyperref} 
\theoremstyle{dgthm}

\theoremstyle{dgdef}

\newcommand{\mr}{\mathbf{r}}

\newcommand{\ptk}[1]{{\color{black}#1}}

\usepackage[resetlabels,labeled]{multibib}

\begin{document}
\title{Confocal polarization tomography of dielectric nanocavities}

\author{F. Schröder}
    \affiliation{Department of Electrical and Photonics Engineering, Technical University of Denmark, Ørsteds Plads 343, 2800 Kgs. Lyngby, Denmark}
    \affiliation{NanoPhoton - Center for Nanophotonics, Technical University of Denmark, Ørsteds Plads 345A, 2800 Kgs. Lyngby, Denmark}
\author{M. P. van Exter}
    \affiliation{Huygens-Kamerlingh Onnes Laboratory, Leiden University, P.O. Box 9504, 2300 RA Leiden, The Netherlands}    
\author{M. Xiong}
    \affiliation{Department of Electrical and Photonics Engineering, Technical University of Denmark, Ørsteds Plads 343, 2800 Kgs. Lyngby, Denmark}
    \affiliation{NanoPhoton - Center for Nanophotonics, Technical University of Denmark, Ørsteds Plads 345A, 2800 Kgs. Lyngby, Denmark}
\author{G. Kountouris}
    \affiliation{Department of Electrical and Photonics Engineering, Technical University of Denmark, Ørsteds Plads 343, 2800 Kgs. Lyngby, Denmark}
    \affiliation{NanoPhoton - Center for Nanophotonics, Technical University of Denmark, Ørsteds Plads 345A, 2800 Kgs. Lyngby, Denmark}
\author{M. Wubs}
    \affiliation{Department of Electrical and Photonics Engineering, Technical University of Denmark, Ørsteds Plads 343, 2800 Kgs. Lyngby, Denmark}
    \affiliation{NanoPhoton - Center for Nanophotonics, Technical University of Denmark, Ørsteds Plads 345A, 2800 Kgs. Lyngby, Denmark}
\author{P. T. Kristensen}
    \affiliation{Department of Electrical and Photonics Engineering, Technical University of Denmark, Ørsteds Plads 343, 2800 Kgs. Lyngby, Denmark}
    \affiliation{NanoPhoton - Center for Nanophotonics, Technical University of Denmark, Ørsteds Plads 345A, 2800 Kgs. Lyngby, Denmark}
\author{N. Stenger}
    \email[Correspondence email address: ]{niste@dtu.dk}
    \affiliation{Department of Electrical and Photonics Engineering, Technical University of Denmark, Ørsteds Plads 343, 2800 Kgs. Lyngby, Denmark}
    \affiliation{NanoPhoton - Center for Nanophotonics, Technical University of Denmark, Ørsteds Plads 345A, 2800 Kgs. Lyngby, Denmark}
    
\date{\today} 

\begin{abstract}
We employ polarization tomography to characterize the modal properties of a dielectric nanocavity with sub-wavelength mode confinement. Our analysis of reflection spectra shows that the Fano-lineshape depends strongly
on the polarization in a confocal configuration, and that 
the lineshape can be transformed into a Lorentzian-like peak for a certain polarization. For this polarization setting, the background is almost fully suppressed in a finite range of frequencies. This enables us to identify another resonance
that has not yet been experimentally reported for these nanocavities.
Lastly, we use symmetry-forbidden polarizations and show that, surprisingly, the modal resonance features
of the system remain visible.
\end{abstract}


\maketitle

\section{Introduction}
\label{sec:introduction}
Optical nanocavities, such as photonic crystal cavities \cite{Painter1999, Akahane2003} and nanobeam cavities \cite{Ota2018}, play a crucial role in enhancing light-matter interactions \cite{Lodahl2015} and  
are vital for applications ranging from efficient nanolasers \cite{Matsuo2010, Crosnier2017} to quantum technologies \cite{Michler2000, Reithmaier2004, Tiecke2014, Vajner2022}. Recent breakthroughs in dielectric cavity design have enabled 
confinement of electromagnetic fields on sub-wavelength scales \cite{Robinson2005, Hu2016, Choi2017, Hu2018, Babar2023,Albrechtsen2022, Xiong2024}. In particular, cavities resulting from inverse design by
topology optimization \cite{Jensen2011,Molesky2018,Wang2018b} have been realized in silicon \cite{Albrechtsen2022} and indium phosphide \cite{Xiong2024} and have inspired further simplified designs \cite{Kountouris2022}. Importantly, these extreme dielectric confinement (EDC) cavities are not limited by absorption in the material and can
achieve quality factors several orders of magnitude higher
than those of plasmonic structures \cite{Wang2006,Naik2013,Khurgin2015}.
The emergence of EDC cavities therefore represents a paradigm shift in device design, in which their strong field enhancements, typically achievable only with plasmonic systems but without the associated losses, open the door to unique light-matter interaction regimes.

To fully harness the potential of EDC cavities, it is essential to analyze their spatial and spectral properties in detail. Polarization-based tomography in a confocal microscope \cite{Bueno2002} is an effective technique for mode identification, which enables 
characterization of the 
associated resonance energies, polarization states, linewidths, and spatial distributions, even though the resolution is limited by the spot size of the focused light beam. This versatile method has been successfully applied to various photonic structures, including photonic crystals \cite{Dimopoulos2022} and vertical-cavity surface-emitting lasers \cite{VanExter1998,Willemsen2000}. Reflection spectra of 
\ptk{EDC cavities} typically have a Fano lineshape \cite{Albrechtsen2022, Xiong2024, Babar2023}, \ptk{resulting} 
from interference of \ptk{a spectrally narrow resonance} 
with a slowly varying background \cite{Fano1961}, also observed in other types of cavities, like photonic crystal cavities \cite{Galli2009a, Bekele2019} and plasmonic 
\ptk{resonators} \cite{Ropers2005, Miroshnichenko2010}. This complicates the identification and quantification of resonance energies and quality factors\ptk{, and understanding} 
and exploiting this interference effect has attracted much interest \cite{DeDood2008, Miroshnichenko2010, Lo2012, Yu2014, Limonov2017, Granchi2024}. Several publications also consider the influence of polarization on the Fano lineshape \cite{Avrutsky2013, Huang2013, Chenari2018, Chang2020, Zhang2024}, and theoretical studies \ptk{have} investigated polarization-dependent lineshape in photonic crystal cavities \cite{Vasco2013}.

In this work, we present an adapted method of polarization tomography to explore the spectral, spatial, and polarization properties of topology optimization-inspired EDC cavities. Our approach goes beyond conventional cross-polarization configurations by incorporating \ptk{variation in the} 
polarization angles. We show that the reflection measurements in a confocal geometry \ptk{depends crucially} 
on the polarization\ptk{, and we exploit this fact to} 
demonstrate that the background can be completely eliminated over specific frequency ranges and for a certain elliptical polarization projection of the detected light. The lineshape transforms into a Lorentzian-like peak\ptk{, and the approach enables the} 
isolation of spectrally close \ptk{resonances} 
that are difficult to distinguish in standard configurations\ptk{. In this way we identify an otherwise hidden feature in the experimental spectrum, which we are able to identify with numerical simulations.}

This general method can be applied to other types of nanocavities and complements scattering-type scanning near-field optical microscopy measurements by providing precise information about cavity polarization properties while eliminating interference effects. \ptk{We find, that the method offers impressive} 
detail in characterizing EDC cavities, which is crucial for optimizing light-matter interactions. This, in turn, enables 
\ptk{design and modeling of future devices exploiting} 
nonlinearities at the single-photon level for quantum information processing \cite{Denning2022} and for the development of low-noise lasers for energy-efficient optical communications \cite{Mork2020}.

\section{Polarization tomography of cavity resonances}
\label{sec:theory}
\ptk{Polarization tomography can be conveniently performed using reflection spectroscopy in a confocal geometry, in which focused input and output beams are used, respectively, to illuminate the sample and collect the scattered light.} The superior spatial control of this \ptk{setup} 
enables a 
selective excitation of specific resonances and a tailored background suppression.

Following the ideas of coupled-mode theory for describing coupled cavity-waveguide systems~\cite{Fan2003}, we take the illumination and collection to be described by the two-dimensional vector functions $\vec{S}_\text{in}(\omega)$ and $\vec{S}_\text{out}(\omega)$. 
They relate to the electric field in the focal plane via $ \vec{E}(x,y,\omega) = \sqrt{\frac{\epsilon_0}{2}}f_\mathrm{c}(x,y)\vec{S}(\omega)$, where $f_\mathrm{c}(x,y)$ describes the electric-field distribution in the focal plane governed by the optical setup \cite{Kristensen2017}. For most setups, $f_\mathrm{c}(x,y)$ is a two-dimensional Gaussian function. Without loss of generality, we scale the vector functions so that $|\vec{S}_\text{in}(\omega)|^2$ and $|\vec{S}_\text{out}(\omega)|^2$ provide the delivered input and collected output power, respectively, and for a chosen focal point of the illumination and collection optics $\mr_0=(X,Y,Z)$ 
we can relate them 
by use of a $2\times2$ reflection matrix $r(\omega)$ as
\begin{equation}
\label{eq:reflection-matrix}
     \begin{pmatrix} S_{\mathrm{out},x}(\omega) \\ S_{\mathrm{out},y}(\omega) \end{pmatrix} =
    \begin{pmatrix} r_{xx}(\omega) & r_{xy}(\omega) \\ r_{yx}(\omega) & r_{yy}(\omega) \end{pmatrix} 
    \begin{pmatrix} S_{\mathrm{in},x}(\omega) \\ S_{\mathrm{in},y}(\omega) \end{pmatrix}.
\end{equation}
In general, $r(\omega)$, and in turn $\vec{S}_\text{out}(\omega)$, depend on the focal point position $\mr_0$ with respect to the cavity, and we exploit this dependence to characterize the spatial properties of the detected signal in Sec. \ref{sec:results}.
If the sample has $x$ and $y$ mirror symmetry, and if the illumination and detection are on-axis, the off-diagonal elements of $r(\omega)$ vanish. 
Practical samples are never perfect, and small deviations from mirror symmetry in sample or optical alignment can create \ptk{small but} non-zero off-diagonal elements. \ptk{Contrary to the off-diagonal elements, the} 
on-diagonal coefficients \ptk{can be relatively large and contain signatures of} 
different resonances that \ptk{typically 
couple primarily} to $x$- or $y$-polarized input \ptk{or} output. 

In optics, sharp spectral resonances are often situated on top of a broad spectral background and are visible as Fano profiles \cite{Fan_03}. For a single resonance, the spectrum of the combined field can then be written as 
\begin{equation}
\label{eq:Fano-vector}
    \vec{S}_{\mathrm{out}}(\omega) = \vec{b}(\omega) + \frac{\vec{a}}{1-\mathrm{i}(\omega-\omega_0)/\gamma} \,,
\end{equation}
where \ptk{$\vec{a}$} is a \ptk{two-dimensional vector} 
related to the field of a discrete mode that resonates at 
frequency $\omega_0$ with a damping rate $\gamma$ and quality factor $Q=\omega_0/(2\gamma)$. \ptk{Mathematically, the discrete modes are known as quasinormal modes~\cite{Ching1998, Kristensen2013, Lalanne2018, Kristensen2020} or resonant states \cite{GarcaCaldern1976, Muljarov2010, Both2021}, and are defined as solutions to the wave equation with suitable radiation conditions to model light propagating away from the resonator.}
The spectral background $\vec{b}(\omega)$ is typically slowly varying. The frequency dependence of $\vec{a}$ can be neglected since it is determined by the frequency dependence of the optical setup, which is expected to be constant in the spectral range of the cavity mode's linewidth. 
The power spectrum produced by the field in Eq. \ref{eq:Fano-vector} has the form of the previously-reported Fano lineshape \cite{Galli2009a, Albrechtsen2022}:
\begin{equation}
    \label{Fano_fit_fun}
    P(\omega) = A_0(\omega) + F_0\dfrac{\left(q + (\omega - \omega_{\mathrm{0}})/\gamma\right)^2}{1 + \left(\omega - \omega_{\mathrm{0}})/\gamma\right)^2}\, ,
\end{equation}
where $F_0$ is related to the amplitude, and $A_0(\omega)$ denotes the offset spectrum, which determines the background via $A_0(\omega)+F_0$. See Sec. \ref{sec:Fano} in the Supplementary Information \cite{SupplementaryInfo} for an explicit calculation of the power spectrum in Eq. \ref{Fano_fit_fun}.
$P(\omega)$ becomes a Lorentzian dip/peak for $q \rightarrow 0, \pm \infty$, respectively.

The polarizations of the resonant contribution and of the background can be quite different. Whereas $\vec{a}$ is typically linearly polarized, the orientation of $\vec{b}(\omega)$ is generally more complicated. Typically, reflection measurements are performed in a \ptk{cross-polarization} configuration \cite{Bekele2019,Galli2009a,Lee2015,Yu2014} with the input and output polarizations oriented at $-45^\circ$ and $45^\circ$ with respect to $\vec{a}$, see the inset of Fig. \ref{Sample} a). 
The polarization-projected background $\tilde{b}_x(\omega) - \tilde{b}_y(\omega)$ observed in this geometry disappears when $\tilde{b}_x(\omega) = \tilde{b}_y(\omega)$, i.e. when the background has no polarization preference. 
For the general case $\tilde{b}_x(\omega) \neq \tilde{b}_y(\omega)$ the spectral background can, in principle, also be removed by simply placing a quarter-wave plate in front of the analyzer and setting both components at convenient angles. 
\ptk{This follows from the fact that $\vec{b}(\omega)$} has a well-defined polarization at any frequency $\omega$. This suppression should \ptk{be possible at any fixed frequency by using optimized polarization settings.} We demonstrate this general background suppression in a finite range of frequencies in Sec. \ref{sec:results}.

\section{Sample and setup}
\label{sec:sample}
\begin{figure}[t!]
         \centering
         \includegraphics[width = \linewidth]{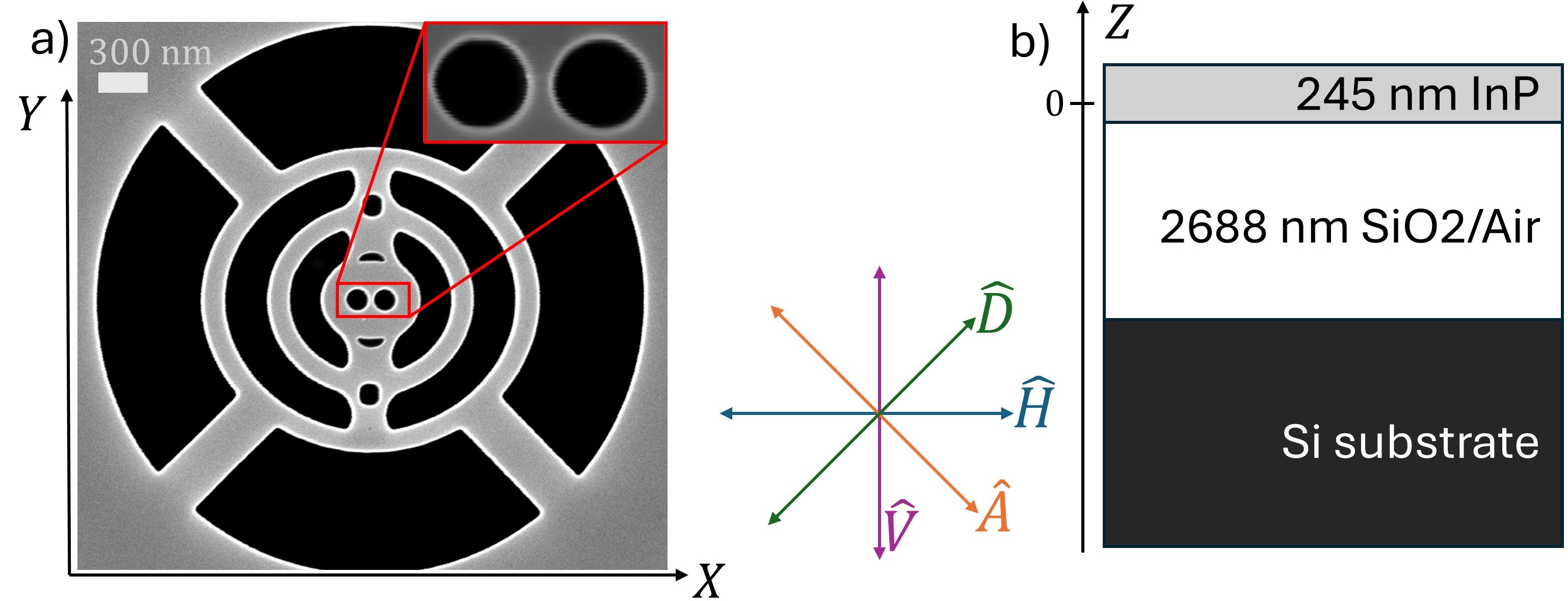}
         \caption{\textbf{a)} SEM image of a nominal equal cavity, defining the cartesian coordinates $X$ and $Y$, as well as the linear polarizations $\hat{H}$, $\hat{V}$, $\hat{D}$ and $\hat{A}$. The center of the cavity is taken as the origin of $X$ and $Y$. \textbf{b)} Sketch of a cross-section of the sample. The SiO$_2$ layer is etched under the cavity region, e.g. in the structured circular region in a) with a diameter of $\approx \SI{3.5}{\micro\metre}$.}
         \label{Sample}
\end{figure}

We investigated an EDC cavity as \ptk{the} one depicted in Fig. \ref{Sample}\ptk{. This design~\cite{Kountouris2022}, which results from a simplification of the cavity in Ref.~\cite{Albrechtsen2022}, features an optical cavity mode with strong field confinement, as verified by scattering-type scanning near-field optical microscopy (s-SNOM). The mode of interest has a nominal resonance energy of $(1.10835\pm2\times10^{-4})$ eV, is linearly polarized along the $\hat{V}$ direction and has a nominal quality factor of $723\pm6$ along with an effective mode volume $0.06\left(\lambda/n_{\mathrm{InP}} \right)^3$ as calculated in the center of the cavity. See Sections~\ref{App_sec_sSNOM} and \ref{App_sec_sim} of the Supplementary Information~\cite{SupplementaryInfo} for details of the s-SNOM measurements and the numerical simulations.} 
\nocite{suppRef1,suppRef2,suppRef3}

The cavity was fabricated in a $\SI{245}{\nano\metre}$ thick layer of indium phosphide (InP) on $\SI{2688}{\nano\metre}$ silicon dioxide (SiO$_2$) on a silicon (Si) substrate as described in \cite{Xiong2024}. The SiO$_2$ layer served as a sacrificial layer, yielding a membranized cavity. \ptk{In the experiments we performed spatial scans}, 
where we moved the sample in the $X-$ and $Y-$directions, as defined in Fig. \ref{Sample} a). Moreover, we pursued reflection measurements for several linear polarizations, horizontal $\hat{H}$, vertical $\hat{V}$, diagonal $\hat{D}$, and antidiagonal $\hat{A}$, cf. Fig. \ref{Sample} a).

\begin{figure}[t!]
         \centering
         \includegraphics[width = \linewidth]{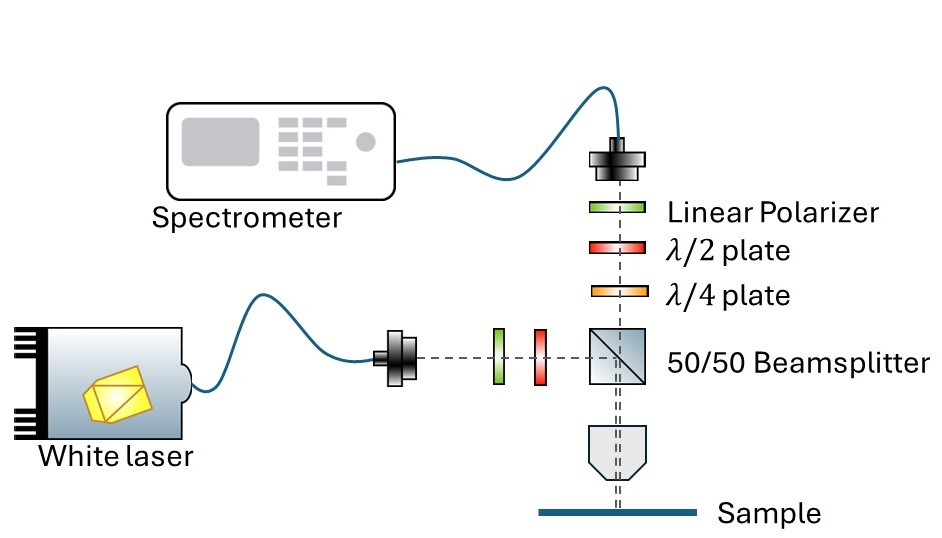}
         \caption{Sketch of the confocal reflection spectroscopy setup. The components are described in the main text.}
         \label{Setup}
\end{figure}

The reflection measurements were carried out in a standard confocal geometry, as depicted in Fig. \ref{Setup}. A supercontinuum laser (SuperK COMPACT, NKT Photonics) was used as a white light source in a spectral range from $\SI{450}{\nano\metre}$ to $\SI{2400}{\nano\metre}$. \ptk{We used a long-pass filter with a cut-on wavelength of $\SI{1100}{\nano\metre}$} 
to prevent saturation of the detector and minimize the fluorescence signal. A linear polarizer and a half-wave ($\lambda/2$) plate \ptk{were used to control the} 
polarization 
of the input \ptk{light, which} 
was focused on the sample with a $50 \times$ microscope objective (LCPLN50XIR, Nikon, NA=0.65). A $50/50$ beamsplitter (BSW29R, Thorlabs) \ptk{enabled illumination and detection} 
through the same objective to measure in a reflection geometry. Another linear polarizer and $\lambda/2$ \ptk{plate-combination} ensured polarization control of the reflected beam. All waveplates were achromatic for broadband operation. For spatial filtering, we collected the reflected beam with a single-mode fiber in the detection path. The collection fiber was coupled to a spectrometer equipped with an indium gallium arsenide camera. All measurements were carried out with a $\SI{150}{\lines/\milli\metre}$ grating. A quarter-wave ($\lambda/4$) was 
\ptk{inserted in certain specific cases, but unless} explicitly stated otherwise, the data presented were acquired without \ptk{the} $\lambda/4$ plate. 

All acquired spectra have been normalized by a reference spectrum taken on a monocrystalline gold (Au) flake of high quality \cite{Casses2022}. 
Additionally, we recorded spectra as a function of the $Z$-position. This way, the out-of-plane confinement of the focal spot can be determined to have a full-width half-maximum $\mathrm{FWHM} = \SI{2.0\pm 0.1}{\micro\metre}$ 
at $E_{\mathrm{ph}} = \SI{1.116}{\eV}$, see Sec. \ref{App_sec_setup} \ptk{of the Supplementary Information~\cite{SupplementaryInfo}.}

The transmission and reflection coefficients of the beamsplitter were polarization- and wavelength-dependent. We have determined these coefficients by comparing spectra measured under different polarization conditions and used this information to convert most of our data into reflectivity spectra, see also Sec. \ref{App_sec_BS} of the Supplementary Information \cite{SupplementaryInfo}.

\begin{figure}[t!]
         \centering
         \includegraphics[width = \linewidth]{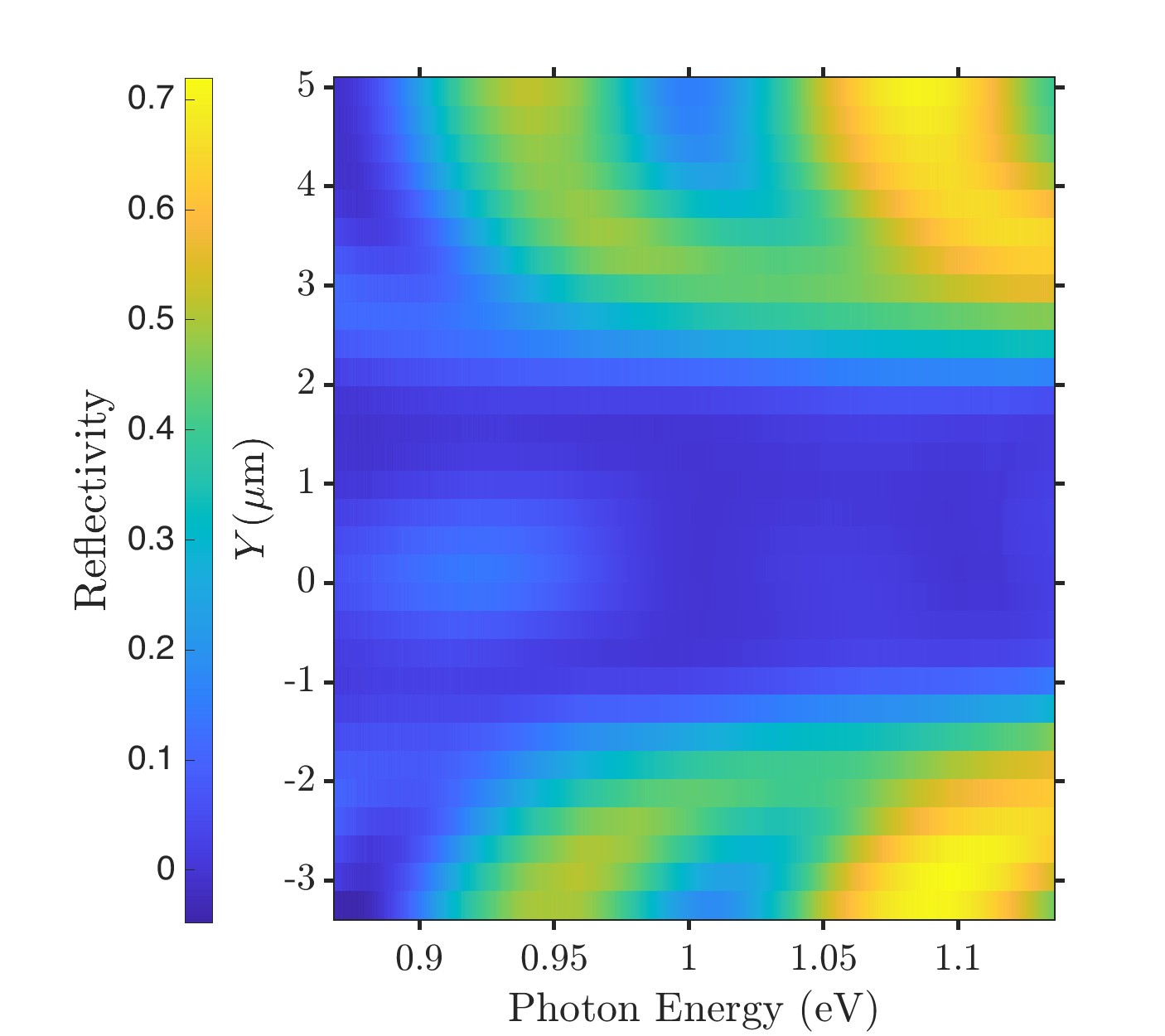}
         \caption{Reflection spectra of the cavity as a function of the $Y$-position at $X = 0$ in parallel polarization ($\hat{E}_{\mathrm{in}} = \hat{E}_{\mathrm{out}} = \hat{D}$).}
         \label{YScan_DD}
\end{figure}
To assess the lateral resolution of our setup, we recorded reflection spectra while moving the sample under the objective with the input- and output polarization parallel to each other. 
Fig. \ref{YScan_DD} shows \ptk{an example of a} 
$Y$-scan with the 
\ptk{input and output electric field-polarizations, $\hat{E}_{\mathrm{in}}$ and $\hat{E}_{\mathrm{out}}$, both} 
along the \ptk{diagonal $\hat{D}$}. 
Clearly, the overall reflection is higher \ptk{from} 
the substrate than the cavity region, which can be explained by Fresnel reflection from the InP/SiO$_2$ substrate. 
Notably, peaks are observed around $\SI{0.94}{\eV}$ and $\SI{1.09}{\eV}$. 
Those are well-explained by Fabry-Pérot type resonances in the SiO$_2$ layer sandwiched between two materials with higher refractive index. 
Indeed, with the resonance condition for a Fabry-Pérot cavity $m \lambda_m = 2n_{\mathrm{SiO2}}d$, where $m$ denotes the order of the resonance, $d$ the thickness and $n_{\mathrm{SiO2}}$ the refractive index of the SiO$_2$ layer, one calculates $E_{\mathrm{ph, m = 6}} = \SI{0.95}{\eV}$ and $E_{\mathrm{ph, m = 7}} = \SI{1.09}{\eV}$.

The reflected signal reduced drastically when the beam was focussed in the void regions next to the outer rings of the cavity and increased slightly when the focal spot was in the center of the cavity \ptk{close to} 
$Y = 0$\ptk{, cf. Fig.~\ref{YScan_DD} and Fig.~\ref{YScan_DD_Sum} of the Supplementary Information~\cite{SupplementaryInfo}}. We determined the spatial resolution of the setup by integrating the spectra and fitting the edges with an error function \ptk{as expected from the} 
convolution of a Gaussian resolution function and an edge function, \ptk{see Fig.~\ref{YScan_DD_Sum} of the Supplementary Information~\cite{SupplementaryInfo}. In this way, we found a} 
FWHM for the lateral resolution of $\SI{1.0\pm 0.2}{\micro \metre}$\ptk{, which is} 
close to the theoretical \ptk{value of} $\SI{0.87}{\micro \metre}$ that would correspond to a diffraction-limited spot at $E_{\mathrm{ph}} = \SI{1.116}{\eV}$. 

\section{Results}
\label{sec:results}
\begin{figure}[t!]
         \centering
         \includegraphics[width = \linewidth]{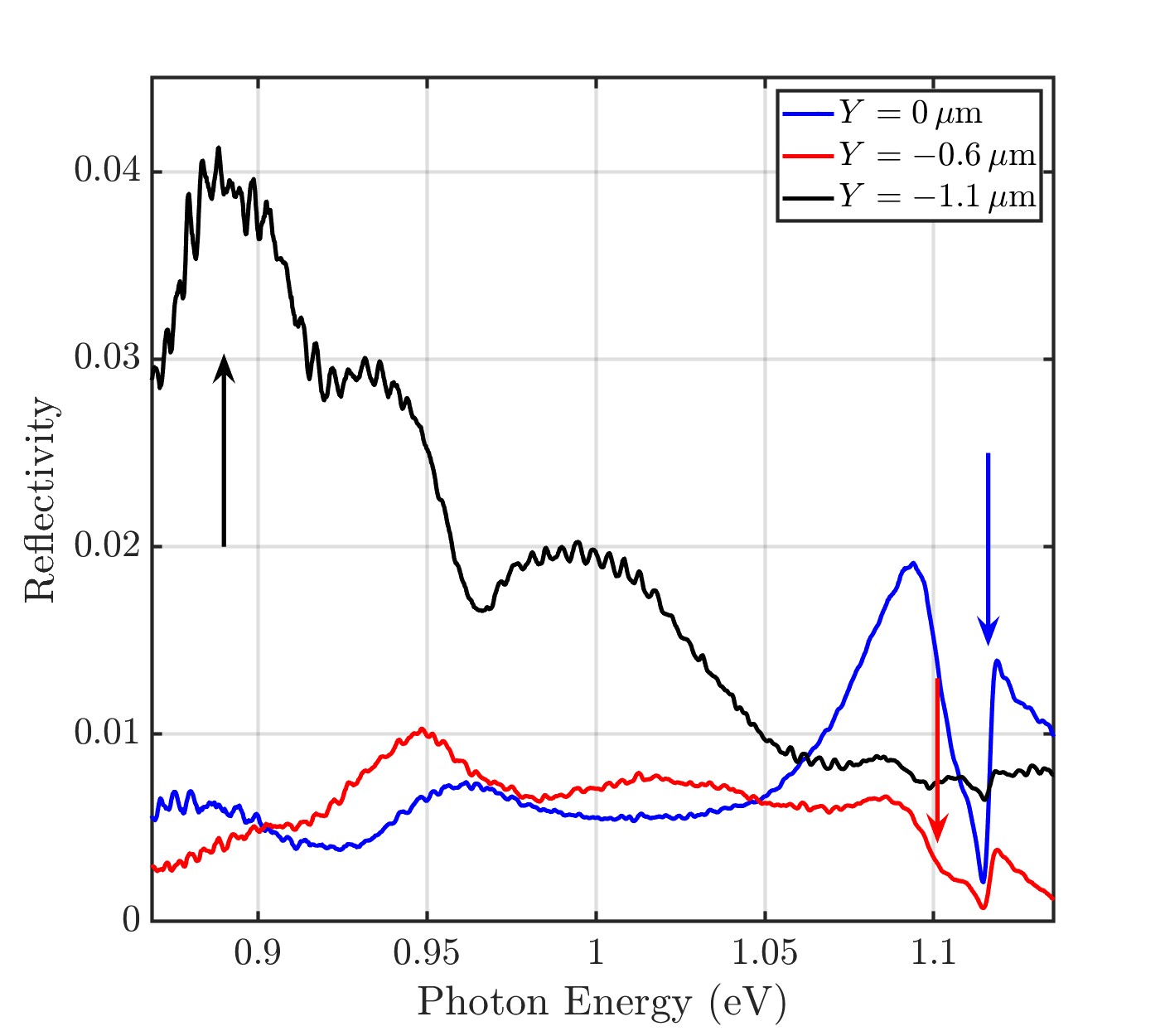}
         \caption{Spectra at different $Y$ positions in the conventional cross-polarization configuration $\hat{E}_{\mathrm{in}} =\hat{A}$, $\hat{E}_{\mathrm{out}} = \hat{D}$. The blue and red arrows mark the resonances of the high-$Q$ mode and of the low-$Q$ mode deduced from fits, respectively (cf. Fig. \ref{FanoFit} and Fig. \ref{FanoFit_lowQ}). The black arrow marks the resonance of the whispering gallery-like mode as a guide to the eye.}
         \label{Cross-pol spectra}
\end{figure}

Fig. \ref{Cross-pol spectra} shows spectra in the standard cross-polarization configuration $\hat{E}_{\mathrm{in}} =\hat{A}$, $\hat{E}_{\mathrm{out}} = \hat{D}$ for different $Y$ positions and for $X = Z = 0$. 
A previously reported cavity mode \cite{Kountouris2022} is visible, marked by the blue arrow. 
As evident from a fit and from simulations (see Fig. \ref{FanoFit} and Fig. \ref{Complex_frequencies} in the Supplementary Information \cite{SupplementaryInfo}), this mode has the highest quality factor.
Henceforth, this mode is referred to as the high-$Q$ mode. When the focus spot is moved by $\SI{0.6}{\micro\metre}$ from the center of the cavity, another mode becomes faintly visible at $E_{\mathrm{ph}} = \SI{1.10}{\eV}$ (see red arrow in Fig. \ref{Cross-pol spectra}).  We deduce a lower quality factor of this mode (see Fig. \ref{FanoFit_lowQ} and Fig. \ref{Complex_frequencies} in the Supplementary Information \cite{SupplementaryInfo}), which is why we label this mode as the low-$Q$ mode. Finally, at about $\SIrange{0.9}{1.2}{\micro\metre}$ away from the center, a third mode appears spectrally detuned from the other two, see also Fig. \ref{Y_scan_crossPol}. 
This could be attributed to a mode forming in the outer rings of the cavity, similar to a whispering gallery mode (see also Fig. \ref{Sim_lowQ} b)). Indeed, the spatial separation of $\Delta Y = \SI{2}{\micro\metre}$, read off from Fig. \ref{Y_scan_crossPol}, fits well with the diameter of the outer rings of the cavity, cf. Fig. \ref{Sample}.

\begin{figure}[t!]
    \centering
    \includegraphics[width=\linewidth]{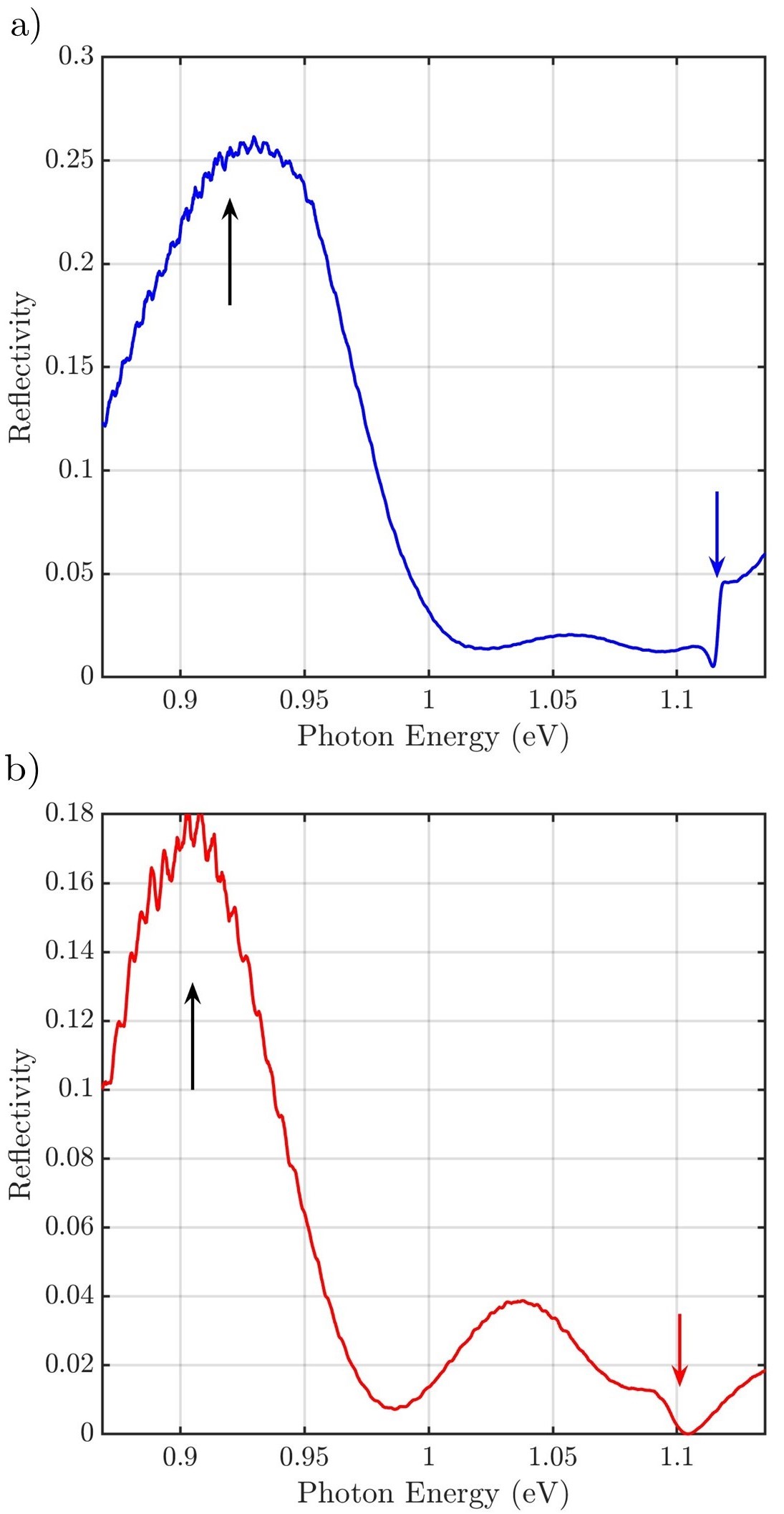}
    \caption{Reflection spectrum in the center of the cavity ($X = Y = Z = 0$), recorded in parallel polarization. The black arrow marks the resonance energy of the Fabry-Perot mode as a guide to the eye. \textbf{a)} $\hat{E}_{\mathrm{in}} =\hat{V}$, $\hat{E}_{\mathrm{out}} = \hat{V}$. The blue arrow marks the resonance energy of the high-$Q$ mode deduced from a fit (cf. Fig. \ref{FanoFit}). \textbf{b)} $\hat{E}_{\mathrm{in}} =\hat{H}$, $\hat{E}_{\mathrm{out}} = \hat{H}$. The red arrow marks the resonance energy of the low-$Q$ mode deduced from a fit (cf. Fig. \ref{FanoFit_lowQ})}
    \label{highQ_lowQ}
\end{figure}

The high-$Q$ mode is polarized along the $\hat{V}$ direction (cf. Fig. \ref{Sim_highQ}) \cite{Kountouris2022}. Therefore, we measure the reflection spectrum in a parallel configuration, where the input and output polarizations are aligned with the polarization of the high-$Q$ mode. Fig. \ref{highQ_lowQ} a) depicts a reflection spectrum for  $\hat{E}_{\mathrm{in}} =\hat{V}$, $\hat{E}_{\mathrm{out}} = \hat{V}$. The high-$Q$ mode, marked by the blue arrow, is clearly visible, while the low-$Q$ mode is suppressed. Moreover, a peak at $E_{\mathrm{Ph}} = \SI{0.93}{\eV}$ can be observed, which is close to a Fabry-Pérot mode $E_{\mathrm{ph, m = 4}} = \SI{0.92}{\eV}$ forming in the air layer ($n_{\mathrm{Air}}$) between the cavity and the silicon substrate.

The reflection spectrum is fitted with Eq. \ref{Fano_fit_fun} close to the resonance of the high-$Q$ mode. For the sake of reducing the number of free-fitting parameters, $A_0(E_{\mathrm{ph}})$ is assumed to be a linear function of $E_{\mathrm{ph}}$ in the vicinity of the resonance. We note that this effectively is a first-order Taylor approximation of the offset spectrum. The fit of the spectrum in the center of the EDC cavity for $\hat{E}_{\mathrm{in}} =\hat{V}$, $\hat{E}_{\mathrm{out}} = \hat{V}$ yields $E_{\mathrm{0, high}Q} = \hbar \omega_{0,\mathrm{high}Q} = \SI{1.1162\pm0.0001}{\eV}$ and $Q_{\mathrm{high}Q} = 265 \pm 8$, cf. Fig. \ref{FanoFit}.

Measuring the reflection with $\hat{E}_{\mathrm{in}} =\hat{H}$, $\hat{E}_{\mathrm{out}} = \hat{H}$, we suppress the reflection of the high-$Q$ mode, while the low-$Q$ mode remains clearly visible (cf. Fig. \ref{highQ_lowQ} b)). We conclude that the low-$Q$ mode is predominantly polarized along the $\hat{H}$ direction. A fit with Eq. \ref{Fano_fit_fun} of the low-$Q$ mode yields $E_{\mathrm{0, low}Q} = \SI{1.1007 \pm 0.0003}{\eV}$ and $Q_{\mathrm{low}Q} = 48 \pm 1$, see Fig. \ref{FanoFit_lowQ}. Consequently, the low-$Q$ mode has a lower quality factor, lower resonance energy, and orthogonal polarization compared to the high-$Q$ mode. An investigation of the spatial extent of those two modes, both with far-field reflection measurements in the confocal geometry as well as near-field measurements with a s-SNOM, can be found in Sec. \ref{App_sec_posscan} and \ref{App_sec_sSNOM} in the Supplementary Information \cite{SupplementaryInfo}. Comparing the spectral and spatial data to FEM simulations (see Sec. \ref{App_sec_sim}), we are able to identify the low-$Q$ mode. See Fig. \ref{Sim_highQ} and \ref{Sim_lowQ} in the Supplementary Information \cite{SupplementaryInfo} for the mode profiles of the high-$Q$ and of the low-$Q$ mode, respectively.


\begin{figure}[t!]
    \centering
    \includegraphics[width=\linewidth]{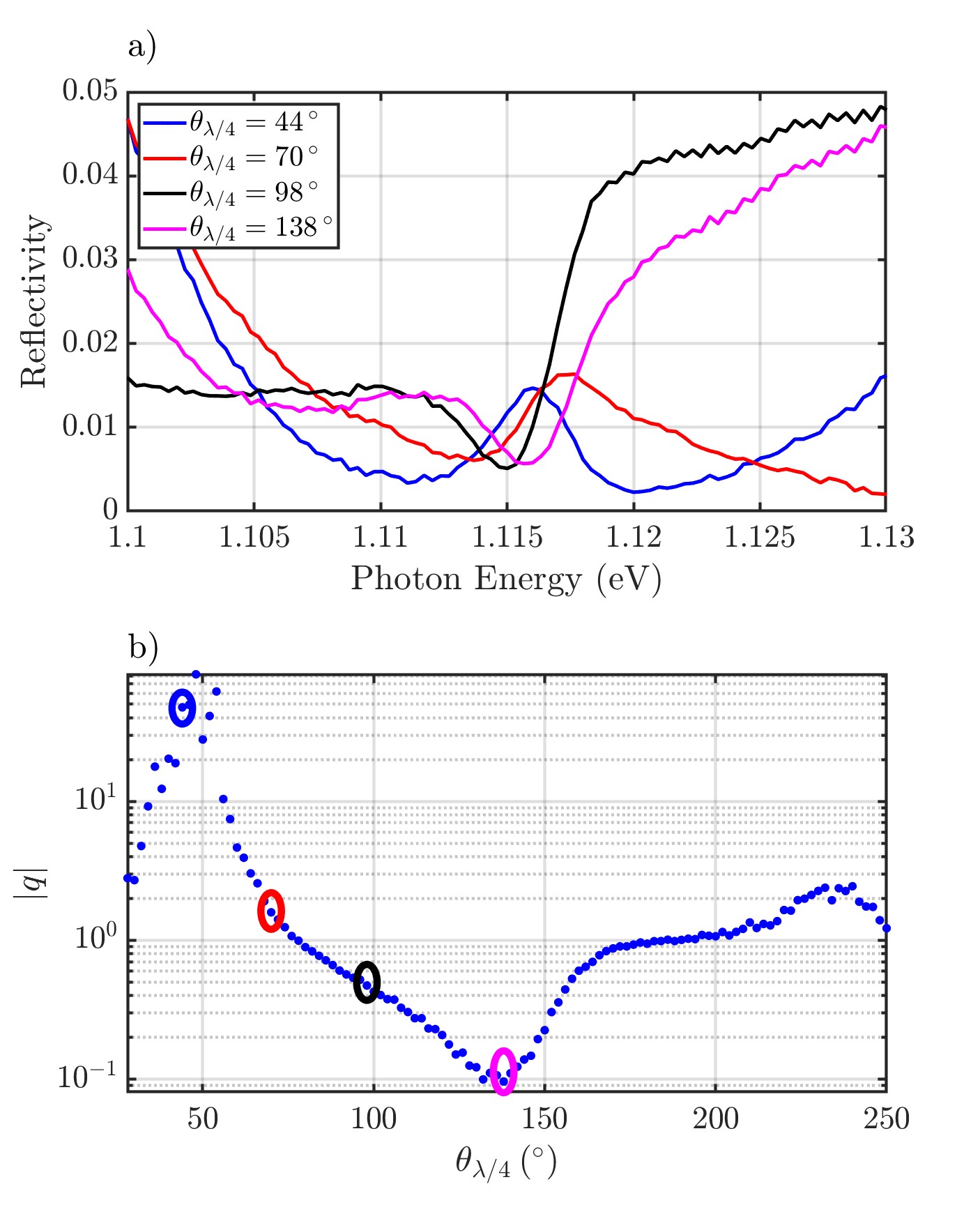}
    \caption{\textbf{a)} Spectra and \textbf{b)} extracted fit parameter $|q|$ for input-polarization $\hat{D}$ and detection polarization rotated by $\theta_{\lambda/2, \mathrm{det}} = \SI{-4}{\degree}$ for various $\theta_{\lambda/4}$ with respect to cross-polarization settings. The values $|q|$ and $\theta_{\lambda/4}$ marked by colored ellipses in b) correspond to the spectra in a) with the same color.}
    \label{Lam4Scan}
\end{figure}

Very importantly, we study the influence of polarization on the lineshape and on the background spectrum and find that the background can be fully suppressed for a finite range of frequencies by detecting a certain polarization.
A $\lambda/4$ plate is now inserted in the detection path to suppress the background completely. For the investigation, the input polarization is fixed to be along the $\hat{D}$ direction, while the $\lambda/2$ and $\lambda/4$ plates in the detection path are rotated by $\theta_{\mathrm{\lambda/2}}$ and $\theta_{\mathrm{\lambda/4}}$, respectively. 
Fig. \ref{Lam4Scan} a) depicts four representative spectra for various settings of $\theta_{\lambda/4}$ and for $\theta_{\mathrm{\lambda/2}} = \SI{-4}{\degree}$ in the spectral region of interest close to the high-$Q$ mode. 
The angles $\theta_{\mathrm{\lambda/2}}$ and $\theta_{\mathrm{\lambda/4}}$ are defined with respect to the standard cross-polarization configuration, such that $\theta_{\mathrm{\lambda/2}} = \theta_{\mathrm{\lambda/4}} = 0$ corresponds to $\hat{E}_{\mathrm{out}} =  \hat{A}$.
Clearly, the spectra are transformed into a Lorentzian-like peak at $\theta_{\lambda/4} = \SIrange{44}{50}{\degree}$.
At that pair of angles ($\theta_{\mathrm{\lambda/2}} = \SI{-4}{\degree}$, $\theta_{\lambda/4} = \SI{44}{\degree}$), the background is almost fully suppressed in a narrow spectral range. 
As the polarization settings differ from simple parallel- or cross-polarization, the detection polarization differs from the optical axis of the beamsplitter and contains even elliptically polarized contributions. Therefore, the calibration of the beamsplitter presented in Sec. \ref{App_sec_BS} is 
challenging, and a correction factor similar to $\chi(\omega)$ as defined in Eq. \ref{correction_factor_BS} in the Supplementary Information \cite{SupplementaryInfo} is difficult to obtain. Therefore, we disregard the correction factor $\chi(\omega)$ and note that 
the reflectivity in Fig. \ref{Lam4Scan} a) might differ
by about a factor of $1.5$. We expect $\chi(\omega)$ to be close to constant in the spectral range depicted in Fig. \ref{Lam4Scan} a), so that the spectral shape is not influenced by $\chi(\omega)$.

To further quantify the effect of the detection polarization on the lineshape, the $\lambda/4$ plate is rotated over a larger range while spectra are recorded. 
Each spectrum is fitted with Eq. \ref{Fano_fit_fun}. 
The parameters $E_0$ and $\gamma$ are fixed by the extracted parameters from the fit of the high-$Q$ mode in parallel polarization (see Fig. \ref{FanoFit}), leaving $F_0$, $q$, and the offset spectrum as free fitting parameters. 
Fig. \ref{Lam4Scan} b) shows the extracted parameter $|q|$ as a function of $\theta_{\lambda/4}$ for $\hat{E}_{\mathrm{in}} = \hat{D}$ and for $\theta_{\lambda/2} = \SI{-4}{\degree}$. 
It can be noted that $q$, and in turn, the detected lineshape, depend crucially on $\theta_{\lambda/4}$. 
Around $\theta_{\lambda/4} = \SI{44}{\degree}-\SI{50}{\degree}$, $q$ diverges, which manifests itself in a Lorentzian-like lineshape, cf. Fig. \ref{Lam4Scan} a). 
Since a Fano lineshape becomes a Lorentzian peak for $q \rightarrow \pm \infty$, the value for $q$ can be subject to a large uncertainty. 
Furthermore, it is apparent from the spectra that the background intensity is almost completely suppressed when $|q|$ diverges. Evaluating the background reflectivity at the resonance energy of the high-$Q$ mode yields $\approx 9\times10^{-5}$ for $\theta_{\lambda/2} = \SI{44}{\degree}$, which is much weaker than $\approx 7\times10^{-3}$ for $\theta_{\lambda/2} = \SI{112}{\degree}$.
In addition, we investigate the influence of the detection polarization on the lineshape without a $\lambda/4$ plate in appendix \ref{App_sec_polser}. As apparent from Fig. \ref{Lam2Scan}, we do not observe a Lorentzian-like lineshape without a $\lambda/4$ plate, demonstrating the important influence of the $\lambda/4$ plate on the lineshape.
It is a main finding of this study that the background contributing to the Fano lineshape is polarized, and that it is possible to suppress the recorded background completely in a narrow spectral range by measuring in a specific polarization.

\begin{figure}[t!]
    \centering
    \includegraphics[width=1\linewidth]{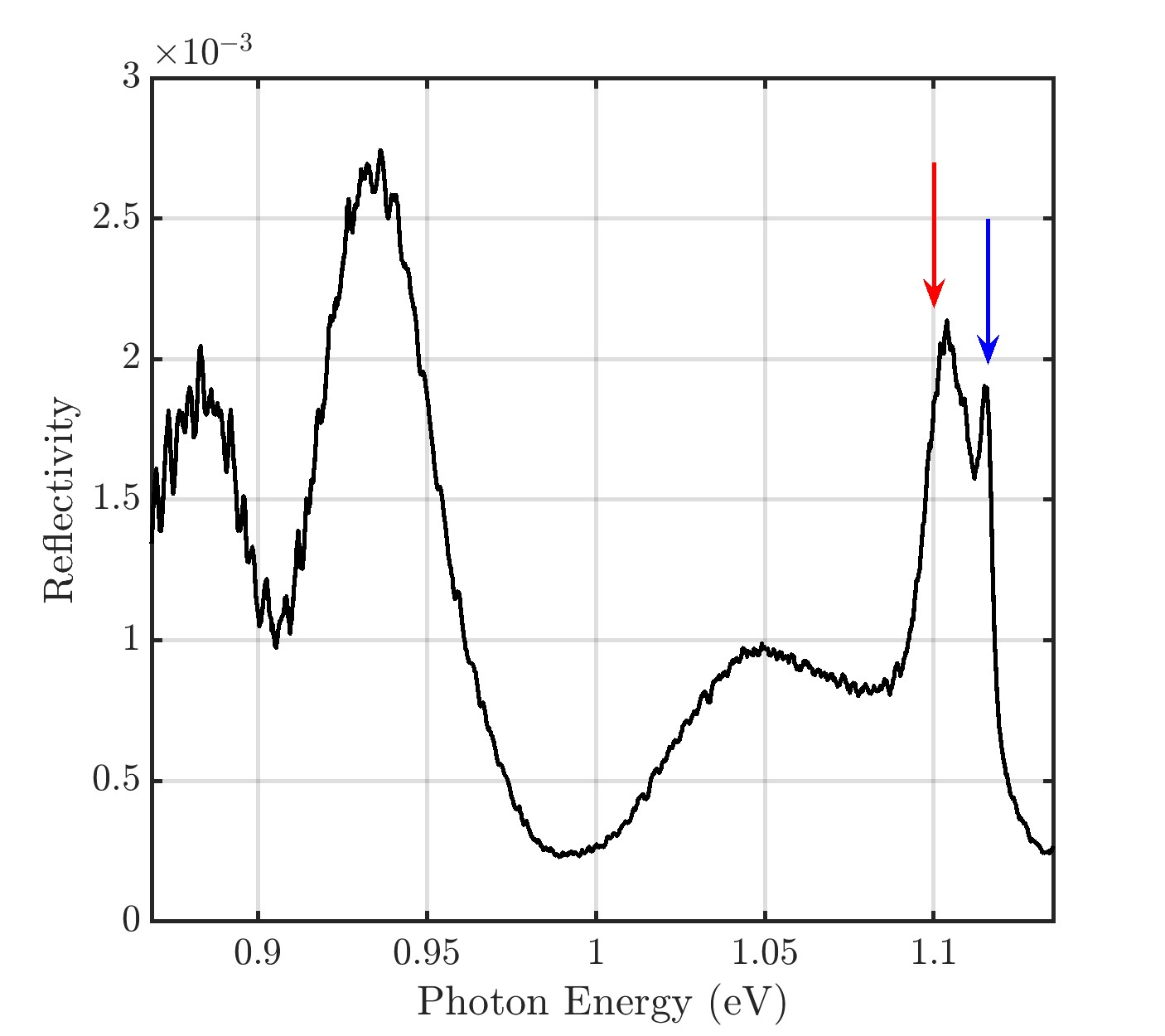}
    \caption{Spectrum in the center of the cavity for $\hat{E}_{\mathrm{in}} = \hat{V}$ and $\hat{E}_{\mathrm{in}} = \hat{H}$. This configuration is symmetry-forbidden. The blue arrow marks the resonance energy of the high-$Q$ mode, the red arrow of the low-$Q$ mode (cf. Fig. \ref{FanoFit} and Fig. \ref{FanoFit_lowQ}).}
    \label{VH_spectrum}
\end{figure}

Lastly, we turn our attention to spectra recorded with $\hat{E}_{\mathrm{in}} = \hat{V}$ and $\hat{E}_{\mathrm{out}} = \hat{H}$. 
In this unusual cross-polarization setting, the input polarization is parallel to the high-$Q$ mode but perpendicular to the low-$Q$ mode, while the detection polarization is parallel to the low-$Q$ mode but perpendicular to the high-$Q$ mode. 
This polarization setting might be considered symmetry-forbidden because it specifically probes the $r_{xy}(\omega)$ and $r_{yx}(\omega)$ elements of the reflection matrix, which are forbidden by symmetry arguments (see Sec. \ref{sec:theory}). 
However, misalignment in the back focal plane and illumination off the center of the cavity can create non-zero off-diagonal elements.
Fig. \ref{VH_spectrum} depicts a spectrum recorded in this configuration in the center of the cavity. 
The symmetry-forbidden nature is visible in the magnitude of the observed reflectivity, which is less than 0.3\% over the full spectral range.
Surprisingly, both the high-$Q$ mode and low-$Q$ mode can still be observed. 
This result demonstrates that choosing these unusual cross-polarization settings allows us to resolve the two individual modes simultaneously.

\section{Concluding discussion}
\label{sec:discussion}
We investigated the spectral and spatial properties of an EDC cavity with polarization tomography. 
Firstly, we formulated a model for the electric field associated with a Fano lineshape intensity similar to the existing literature on this topic \cite{Vasco2013, Avrutsky2013}, considering the vectorial properties of the individual contributions, namely the background and the high-$Q$ mode. This vectorial description of the Fano lineshape is valid in a confocal geometry, though it is not limited to dielectric nanocavities and could be applied to other systems, such as plasmonic nanocavities.

We pursued reflection measurements in a confocal geometry and experimentally observed the previously reported high-$Q$ mode of the EDC cavity, along with a low-$Q$ mode with a lower quality factor, lower resonance energy, and orthogonal polarization compared to the high-$Q$ mode. Moreover, we observe signatures of whispering gallery-like and Fabry-Pérot modes.
We study the modes systematically with polarization tomography. Comparing our results to numerical calculations, we are able to identify the low-$Q$ mode. 
Moreover, we demonstrated that the background, and in turn, the lineshape, crucially depend on the polarization. This very important finding allowed for complete suppression of the background for a certain pair of polarization angles. We showed that the lineshape can be transformed into a Lorentzian-like peak. 
Lastly, we showed that by choosing symmetry-forbidden cross-polarization settings, we can immediately retrieve information about several modes in the system. 
The presented results will be beneficial for the investigation and understanding of reflection measurements of dielectric nanocavities. The method has been proven effective for studying EDC cavities and could be applied to other nanocavities.

\begin{acknowledgements}
We thank Benjamin F. Gøtzsche for fruitful discussions about eigenmode simulations.
\end{acknowledgements}

\section*{Research funding}
This work was supported by the \href{http://dx.doi.org/10.13039/501100001732}{\underline{Danish National Research Foundation}} through NanoPhoton - Center for Nanophotonics, grant number DNRF147. We thank the Otto Moensted Foundation for supporting the Otto Moensted Visiting Professorship of M.P.v.E at the Technical University of Denmark. N. S. thanks the Novo Nordisk Foundation NERD Programme (project QuDec NNF23OC0082957)

\section*{Author contributions}
F.S. and M.P.v.E. conducted
the reflection measurements. M.P.v.E. and P.T.K formulated the theory. F.S. carried out near-field measurements. F.S. and M.P.v.E. analyzed the data. F.S. and G.K. carried out eigenmode simulations. M.X. fabricated the sample. F.S. took SEM images. F.S. and M.P.v.E. wrote the paper. P.T.K., M.W., and N.S. supervised F.S. All authors have given feedback on the manuscript and have accepted responsibility for the entire content of this manuscript.

\section*{Conflict of interest}
Authors state no conflict of interest.


\counterwithin{figure}{section}
\setcounter{figure}{0}
\label{sec:Appendix-modes}

\section{Fit of the lineshapes of the modes}
\label{App_sec_fits}

\begin{figure}[h!]
\vspace*{0in}
         \centering
         \includegraphics[width = 0.9\linewidth]{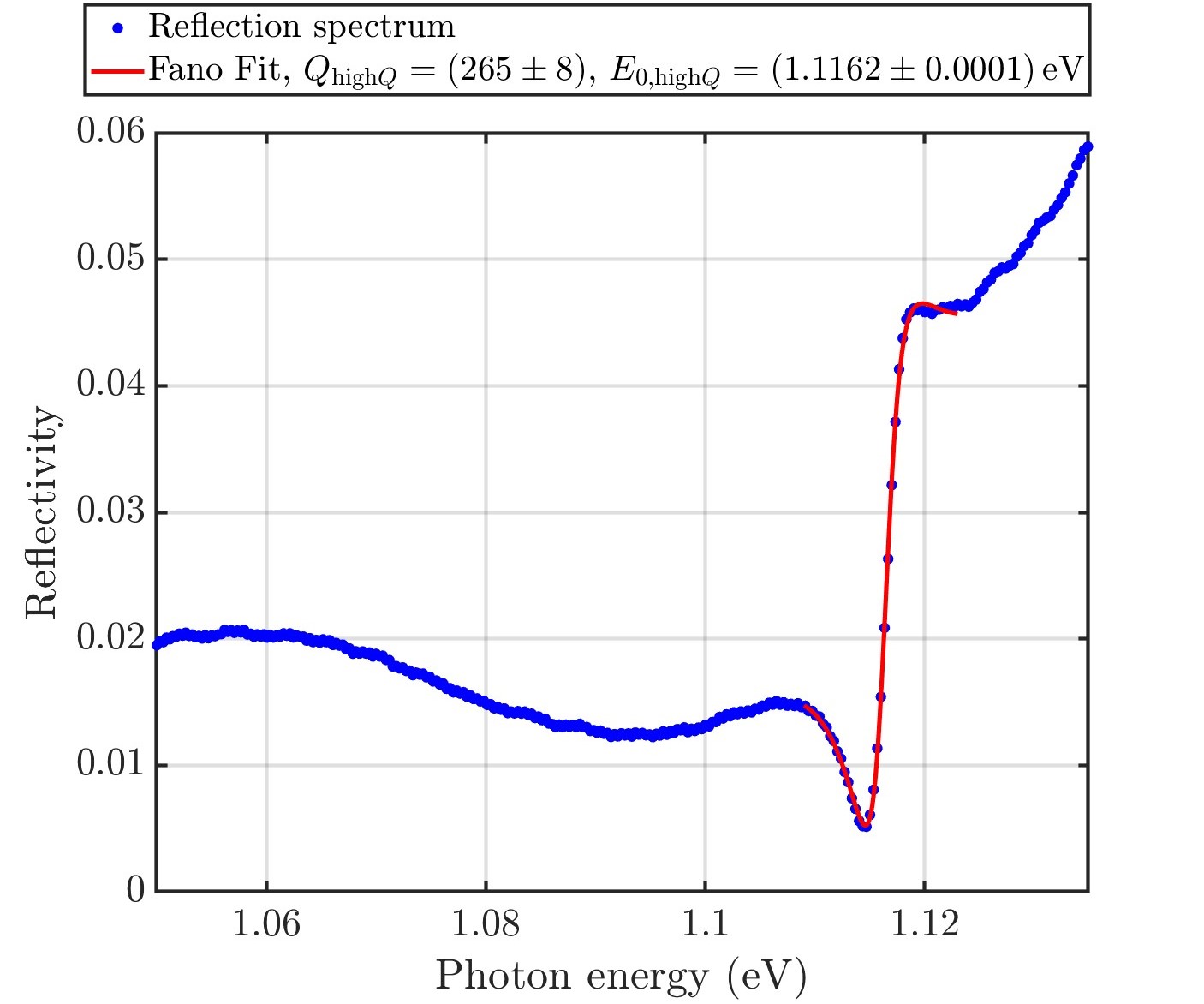}
         \caption{Reflection spectrum in the center of the cavity in parallel-polarization along the polarization of the high-$Q$ mode $\hat{E}_{\mathrm{in}} = \hat{E}_{\mathrm{out}} = \hat{V}$. The high-$Q$ mode has been fitted after Eq. \ref{Fano_fit_fun}.}
         \label{FanoFit}
\end{figure}

\begin{figure}[h!]
\vspace*{0in}
         \centering
         \includegraphics[width = 0.9\linewidth]{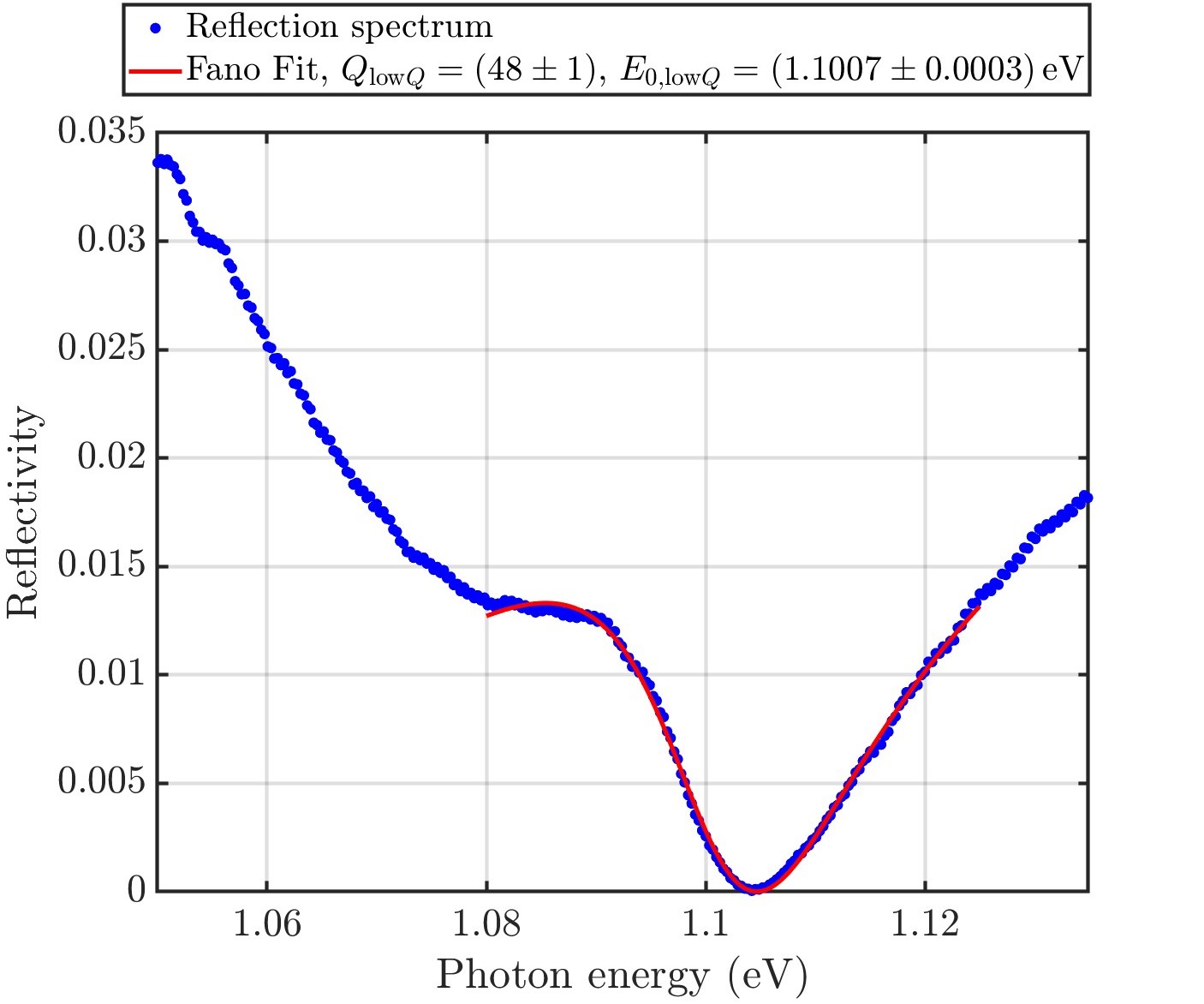}
         \caption{Reflection spectrum in the center of the cavity in parallel-polarization along the polarization of the low-$Q$ mode $\hat{E}_{\mathrm{in}} = \hat{E}_{\mathrm{out}} = \hat{H}$. The low-$Q$ mode has been fitted after Eq. \ref{Fano_fit_fun}.}
         \label{FanoFit_lowQ}
\end{figure} 

Fig. \ref{FanoFit} depicts the reflected spectrum in parallel polarization $\hat{E}_{\mathrm{in}} = \hat{E}_{\mathrm{out}} = \hat{V}$ aligned with the polarization of the high-$Q$ mode at $X = Y = Z = 0$ (cf. Fig. \ref{highQ_lowQ} a)). Moreover, a fit of the high-$Q$ mode with a Fano function as per Eq. \ref{Fano_fit_fun} is shown. The fit yields $E_{\mathrm{0, high}Q} = \SI{1.1162\pm0.0001}{\eV}$ and $Q_{\mathrm{high}Q} = 265 \pm 8$, $F_{0, \mathrm{high}Q} = (14.3\pm0.4) \times 10^{-3}$ and $q_{\mathrm{high}Q} = 0.74 \pm 0.02$. The offset spectrum is determined as $A_0(E_{\mathrm{ph}}) \approx -0.91 + 0.82E_{\mathrm{ph}}/\mathrm{eV}$.

Fig. \ref{FanoFit_lowQ} depicts the reflected spectrum in parallel-polarization $\hat{E}_{\mathrm{in}} = \hat{E}_{\mathrm{out}} = \hat{H}$ aligned with the polarization of the low-$Q$ mode at $X = Y = Z = 0$ (cf. Fig. \ref{highQ_lowQ} b)). A fit of the low-$Q$ mode with a Fano function after Eq. \ref{Fano_fit_fun} yields $E_{\mathrm{0, low}Q} = \SI{1.1007\pm0.0003}{\eV}$ and $Q_{\mathrm{low}Q} = 48 \pm 1$, $F_{0,\mathrm{low}Q} = (16.0\pm0.3) \times 10^{-3}$ and $q_{\mathrm{low}Q} = -0.45 \pm 0.04$. The offset spectrum is determined as $A_0(E_{\mathrm{ph}}) \approx -0.28 + 0.25E_{\mathrm{ph}}/\mathrm{eV}$.

\section{Polarization series without $\lambda/4$ plate}
\label{App_sec_polser}

\begin{figure}[htb!]
\vspace*{0in}
    \centering
    \includegraphics[width=\linewidth]{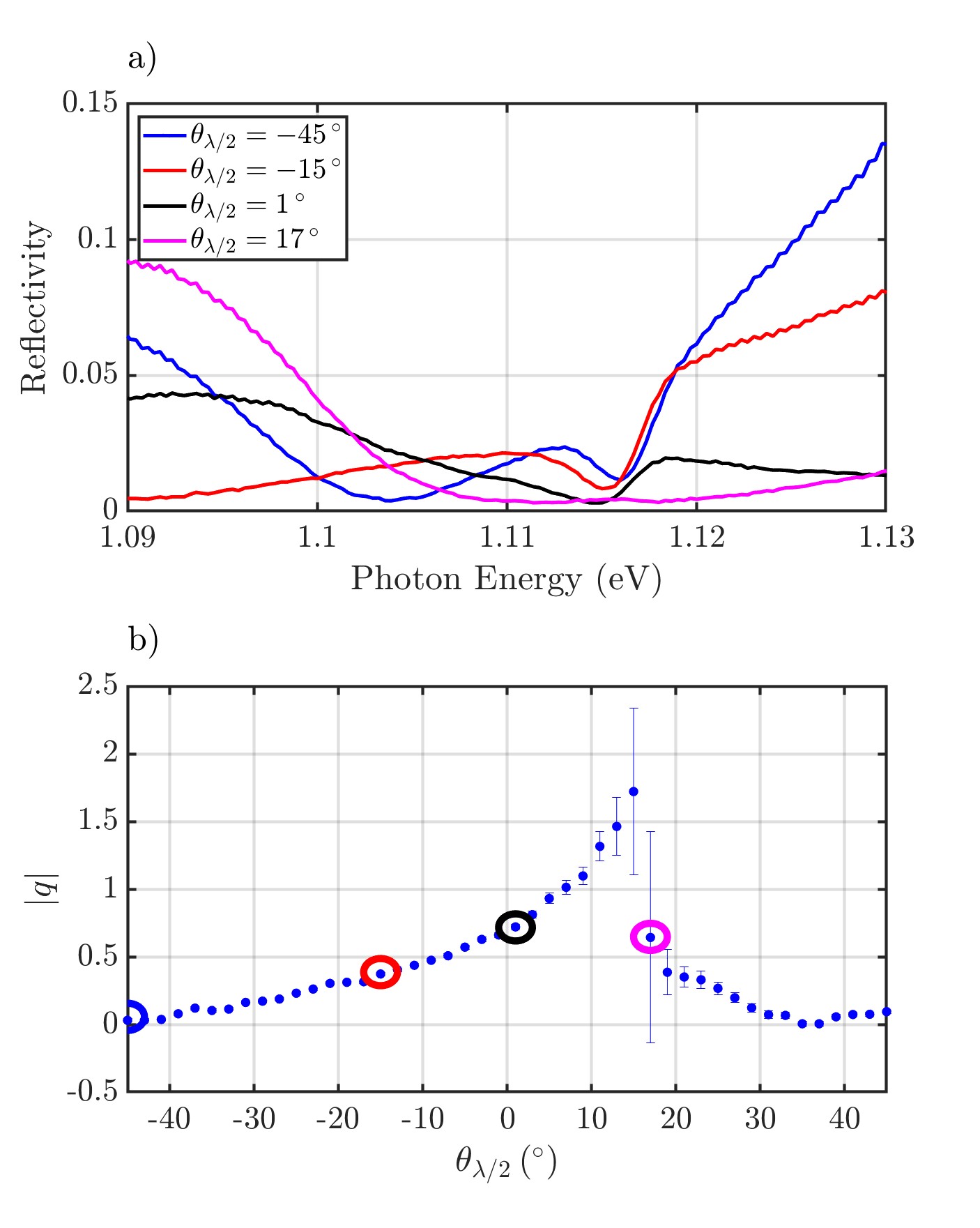}
    \caption{\textbf{a)} Spectra and \textbf{b)} extracted fit parameter $|q|$ for input-polarization $\hat{D}$ and detection polarization rotated by $\theta_{\lambda/2, \mathrm{det}}$ with respect to cross-polarization settings. The values $|q|$ and $\theta_{\lambda/2}$ marked by colored ellipses in b) correspond to the spectra in a) with the same colors.}
    \label{Lam2Scan}
\end{figure}

In addition to the investigation of the lineshape with a $\lambda/4$ plate, we carry out polarization-resolved measurements as a function of $\theta_{\mathrm{\lambda/2}}$. In combination with the analyzer, we effectively rotate the detection polarization by $2 \theta_{\mathrm{\lambda/2}}$. Fig. \ref{Lam2Scan} a) depicts representative reflection spectra for $\hat{E}_{\mathrm{in}} = \hat{D}$ as a function of $\theta_{\mathrm{\lambda/2}}$ and without a $\lambda/4$ plate. It can be seen that when the background vanishes around $\theta_{\mathrm{\lambda/2}} = \SI{15}{\degree}$, also the signal of the high-$Q$ mode vanishes completely. The individual spectra recorded are fitted with Eq. \ref{Fano_fit_fun}, and the extracted fit parameter $|q|$ is shown as a function of $\theta_{\mathrm{\lambda/2}}$ in Fig. \ref{Lam2Scan} b). It is evident that $q$ does not diverge, as in the case when we insert and rotate a $\lambda/4$ plate (cf. Fig. \ref{Lam4Scan}).

\newpage
\
\newpage

\pagebreak
\widetext
\begin{center}
\textbf{\large Supplementary Information: Confocal polarization tomography of dielectric nanocavities}
\end{center}
\setcounter{equation}{0}
\setcounter{section}{0}
\setcounter{figure}{0}
\setcounter{table}{0}
\setcounter{enumiv}{0}
\makeatletter
\renewcommand{\theequation}{S\arabic{equation}}
\renewcommand{\thesection}{S\arabic{section}}
\renewcommand{\thefigure}{S\arabic{figure}}

\section{Explicit calculation of the intensity spectrum} 
\label{sec:Fano}
In this section, we explicitly show that the vector field in Eq. \ref{eq:Fano-vector} is associated with a Fano-lineshape power spectrum (Eq. \ref{Fano_fit_fun}). We start from Eq. \ref{eq:Fano-vector}:
\begin{eqnarray}
    \vec{S}(x) = \vec{b} + \dfrac{\vec{a}}{1-ix} \, ,
\end{eqnarray}
with $x = (\omega-\omega_0)/\gamma$. As described in Sec. \ref{sec:theory}, $\vec{b}$ depends on $\omega$, whereas the frequency-dependence of $\vec{a}$ is neglected. Owing to the linearity of the system, we can choose the phase of the incoming light so that $\vec{a}$ is real. To calculate the power spectrum produced by this field, we take the absolute square \cite{Kristensen2017}:
\begin{eqnarray}
\label{eq:FanoIntenVec}
    P(x) & = & |\vec{S}(x)|^2 = \left(\vec{b} + \dfrac{\vec{a}}{1-ix}\right)\cdot\left(\vec{b}^* + \dfrac{\vec{a}}{1+ix}\right) \nonumber \\
    & = & |\vec{b}|^2 + \dfrac{|\vec{a}|^2}{1+x^2} + \dfrac{\vec{a} \cdot \vec{b}}{1+ix} + \dfrac{\vec{a} \cdot \vec{b}^*}{1-ix} \nonumber \\
    & = & |\vec{b}|^2 + \dfrac{|\vec{a}|^2}{1+x^2} + \dfrac{\vec{a} \cdot \vec{b}(1-ix) + \vec{a} \cdot \vec{b}^*(1+ix)}{1+x^2} \nonumber \\
    & = & |\vec{b}|^2 + \dfrac{|\vec{a}|^2}{1+x^2} + \dfrac{\vec{a} \cdot (\vec{b} +\vec{b}^*)}{1+x^2} + \dfrac{ix\vec{a}\cdot ( \vec{b}^*- \vec{b})}{(1+x^2)} \nonumber \\
    & = & |\vec{b}|^2 + \dfrac{|\vec{a}|^2 + 2\mathfrak{R}{(\vec{a} \cdot \vec{b})} + x2\mathfrak{I}{(\vec{a} \cdot \vec{b})}}{1+x^2} \nonumber \\
    & = & \tilde{B} + \dfrac{\tilde{A} + \tilde{C}x}{1+x^2} \, ,
\end{eqnarray}
where we have introduced
\begin{eqnarray}
\label{eq:coefficientsVec}
\tilde{A} & = & |\vec{a}|^2 + 2\mathfrak{R}{(\vec{a} \cdot \vec{b})} \,, \\
\tilde{B} & = & |\vec{b}|^2 \nonumber \,, \\
\tilde{C} & = & 2\mathfrak{I}{(\vec{a} \cdot \vec{b})} \,,
\end{eqnarray}
in wich $\mathfrak{R}()$ and $\mathfrak{I}()$ take the real and imaginary part of a complex number. 
Now, we can show that this is indeed the Fano lineshape from Eq. \ref{Fano_fit_fun}:
\begin{eqnarray}
\label{eq:FanoInten2}
    P(\omega) & = & A_0 + F_0 \frac{(q+\left(\omega-\omega_0\right)/\gamma)^2}{1+\left(\omega-\omega_0\right)^2/\gamma^2} \nonumber \\
    \Leftrightarrow P(x) & = & A_0 + F_0 \frac{(q+x)^2}{1+x^2} \nonumber \\
    & = & A_0 + F_0 + F_0 \frac{(q+x)^2}{1+x^2} - F_0 \frac{1+x^2}{1+x^2} \nonumber \\
    & = & A_0 + F_0 + F_0 \frac{(q+x)^2 - (1+x^2)}{1+x^2} \nonumber \\
    & = & A_0 + F_0 + F_0 \frac{(q^2-1) + 2qx}{1+x^2} \, ,
\end{eqnarray}
which is exactly the last line of Eq. \ref{eq:FanoIntenVec} with 
\begin{eqnarray}
\label{eq:coefficients}
\tilde{A} & = & F_0 (q^2-1) \nonumber \,, \\
\tilde{B} & = & A_0+F_0 \nonumber \,, \\
\tilde{C} & = & 2F_0q \,.
\end{eqnarray}

\begin{figure}[htb!]
    \centering
    \includegraphics[width = 0.6\linewidth]{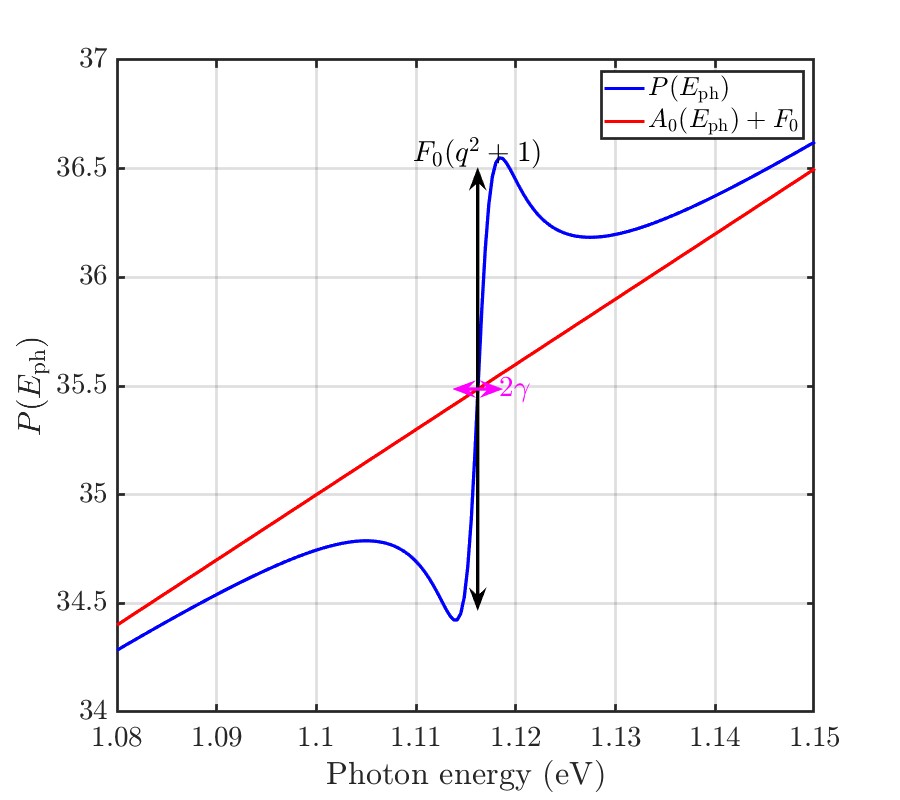}
    \caption{A Fano lineshape for $E_0 = \SI{1.1162}{\eV}$, $\gamma = \SI{2.1}{\milli\eV}$, $F_0 = 1$, $q = 1$ and $A_0(E_{\mathrm{ph}}) = 1 + 30E_{\mathrm{Ph}}/\mathrm{eV}$.}
    \label{FanoLineshapePlot}
\end{figure}

The parameter $A_0 = A_0(\omega)$ is a shorthand notation for the offset spectrum, and the spectral baseline, or background spectrum, is found to be $A_0 + F_0$, see Fig. \ref{FanoLineshapePlot}.
The parameter $F_0$ relates to the amplitude via $F_0(q^2+1)$, see Fig. \ref{FanoLineshapePlot}.
The parameter $q$ determines the overall shape and asymmetry of the spectrum.
\newpage

\section{Reference measurement on Au flake and longitudinal resolution}
\label{App_sec_setup}
\begin{figure}[htb!]
         \centering
         \includegraphics[width = 0.6\linewidth]{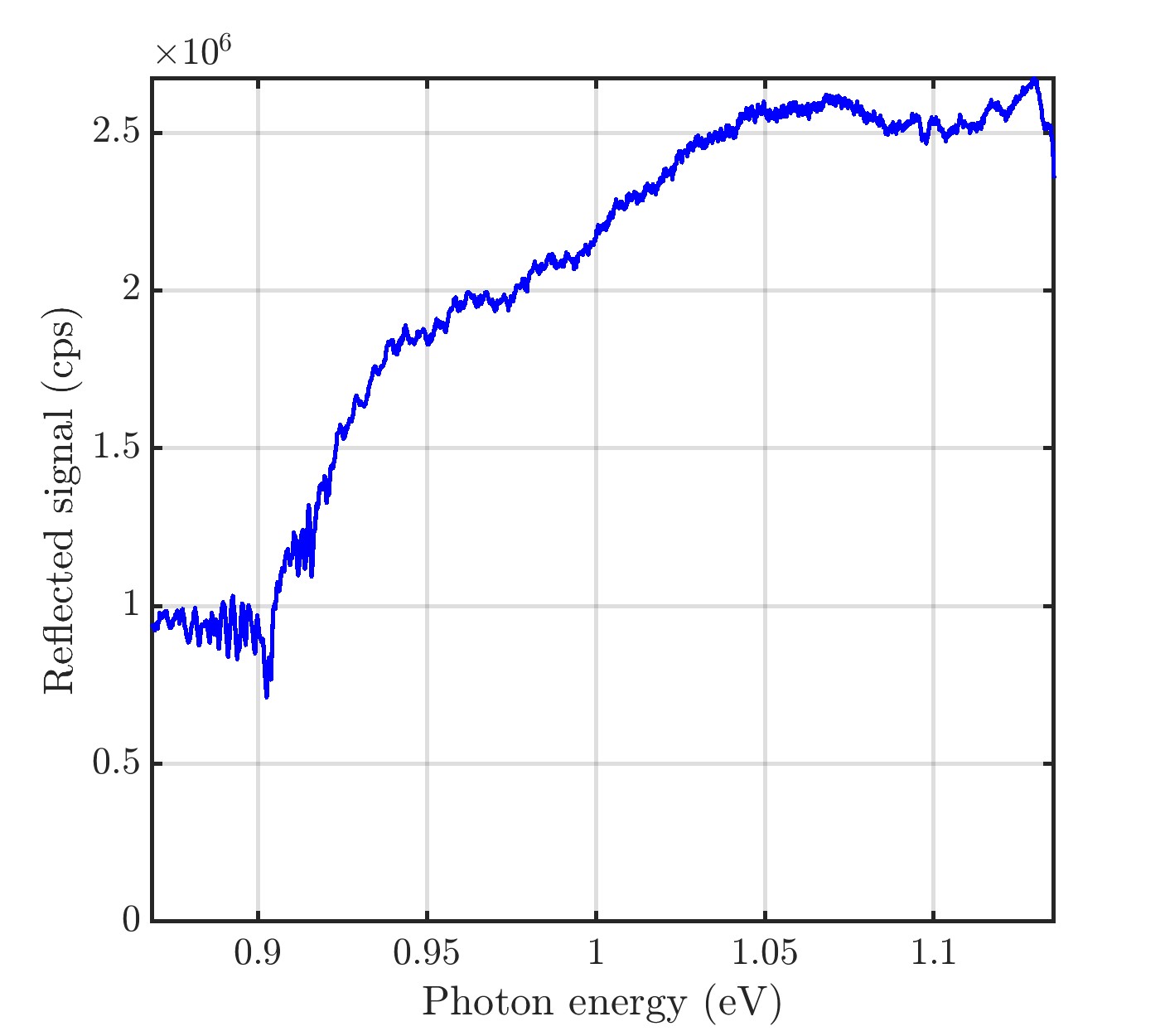}
         \caption{Typical reference spectrum on an Au flake in parallel polarization ($\hat{E}_{\mathrm{in}} = \hat{E}_{\mathrm{out}} = \hat{D}$).}
         \label{Typ_Ref_Spec}
\end{figure}

\begin{figure}[t!]
         \centering
         \includegraphics[width = 0.6\linewidth]{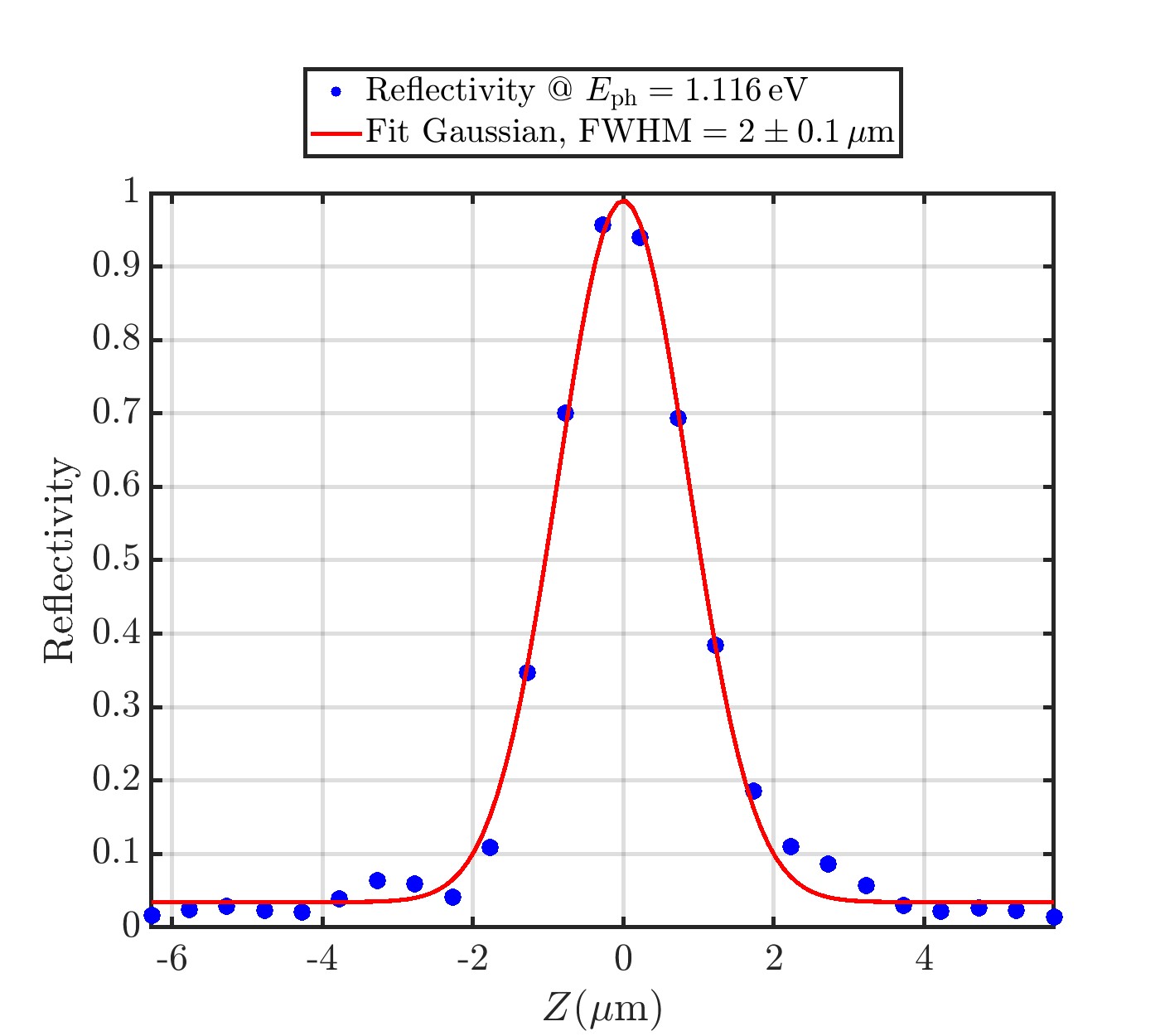}
         \caption{Reflectivity of a gold flake used for reference at $E_{\mathrm{ph}} = \SI{1.116}{\eV}$ as a function of the $Z$-position in parallel polarization ($\hat{E}_{\mathrm{in}} = \hat{E}_{\mathrm{out}} = \hat{D}$).}
         \label{ZYScan_DD_Int}
\end{figure}

A typical reflection spectrum on an Au flake to be used for normalization can be found in Fig. \ref{Typ_Ref_Spec}. Moreover, we measure the reflectivity of the reference gold flake as a function of the $Z$-position to estimate the out-of-plane resolution of the optical setup. Fig. \ref{ZYScan_DD_Int} shows the reflectivity at the photon energy of interest $E_{\mathrm{ph}} = \SI{1.116}{\eV}$, which is the resonance energy of the high-$Q$ cavity mode (see Sec. \ref{App_sec_fits}). Fitting the reflectivity with a Gaussian function yields an out-of-plane FWHM of $\SI{2.0 \pm 0.1}{\micro\metre}$. The focus and the focal depth can be slightly wavelength-dependent due to chromatic aberration.

\newpage
\
\newpage
\section{Calibration of setup and beamsplitter}
\label{App_sec_BS}

All reflection spectra in the main text are presented as reflectivities. 
We obtain these data by comparing the reflection spectrum of our nanocavity sample with that of a gold flake, which acts as a close-to-perfectly-reflecting reference.
Hence, we determine the reflectivity $\tilde{R}$ as
\begin{equation}
    \tilde{R}(\omega) = \dfrac{R_{\mathrm{sam}}(\omega)}{R_{\mathrm{ref}}(\omega)} \,,
\end{equation}
where $R_{\mathrm{sam}}$ and $R_{\mathrm{ref}}$ denote the reflection coefficient of the sample and of the reference, respectively. 
This calibration works fine for measurements with parallel polarization but needs a correction in crossed polarization with an unbalanced beamsplitter.
We will denote the reflection and transmission coefficients of the beamsplitter as $R_s, R_p, T_s$ and $T_p$ for $s$- and $p$-polarized light. 
Measurements on the gold flake with identical input and output polarizations gave almost identical results for $s$ and $p$ polarizations and thereby showed that $R_s T_s = R_p T_p$.
Hence, we only need to introduce a single correction function
\begin{equation}
    \chi(\omega) \equiv \dfrac{R_s(\omega)}{R_p(\omega)} = \dfrac{T_p(\omega)}{T_s(\omega)} \,.
\end{equation}

Consider a setup where the input is $p$-polarized and the detection is $s$-polarized and compare the measured reflection spectrum $R_{\mathrm{sam},ps}(\omega)$ with the $R_{\mathrm{ref},pp}(\omega)$ reference spectrum obtained with $p$-polarized input and output.
The ratio of these spectra yields the measured reflectivity
\begin{equation}
    R_{\mathrm{meas.},ps}(\omega) = \dfrac{R_{\mathrm{sam},ps}}{R_{\mathrm{ref},pp}} = \dfrac{R_s}{R_p} \tilde{R}_{ps}(\omega) =  \chi(\omega) \tilde{R}_{ps}(\omega) \,,
\end{equation}
where $\tilde{R}_{ps}(\omega)$ is the reflectivity. 
A similar calculation for $s$-polarized input and $p$-polarized output yields $R_{\mathrm{meas.},sp}(\omega) = \tilde{R}_{sp}(\omega)/\chi(\omega)$.
To determine $\chi(\omega)$, we divide these two equations, using $\tilde{R}_{ps}(\omega) = \tilde{R}_{sp}(\omega)$ from reciprocity, to find:
\begin{equation}
    \chi(\omega) = \sqrt{\dfrac{\tilde{R}_{\mathrm{meas.},ps}}{\tilde{R}_{\mathrm{meas.},sp}}} \,.
    \label{correction_factor_BS}
\end{equation}

These calculations only work if the input- and output light are aligned with the $s$ and $p$ axes of the beamsplitter. 
We therefore chose to record the $\hat{D}$ and $\hat{A}$ polarizations (cf. Fig \ref{Sample}) with the sample being rotated by $\SI{45}{\degree}$.
For more general polarizations, we would need to include a potential phase shift between $s$- and $p$-polarized light throughout the setup.

\newpage
\section{Lateral resolution of the setup}
\label{App_sec_latres}
\begin{figure}[ht!]
         \centering
         \includegraphics[width = 0.6\linewidth]{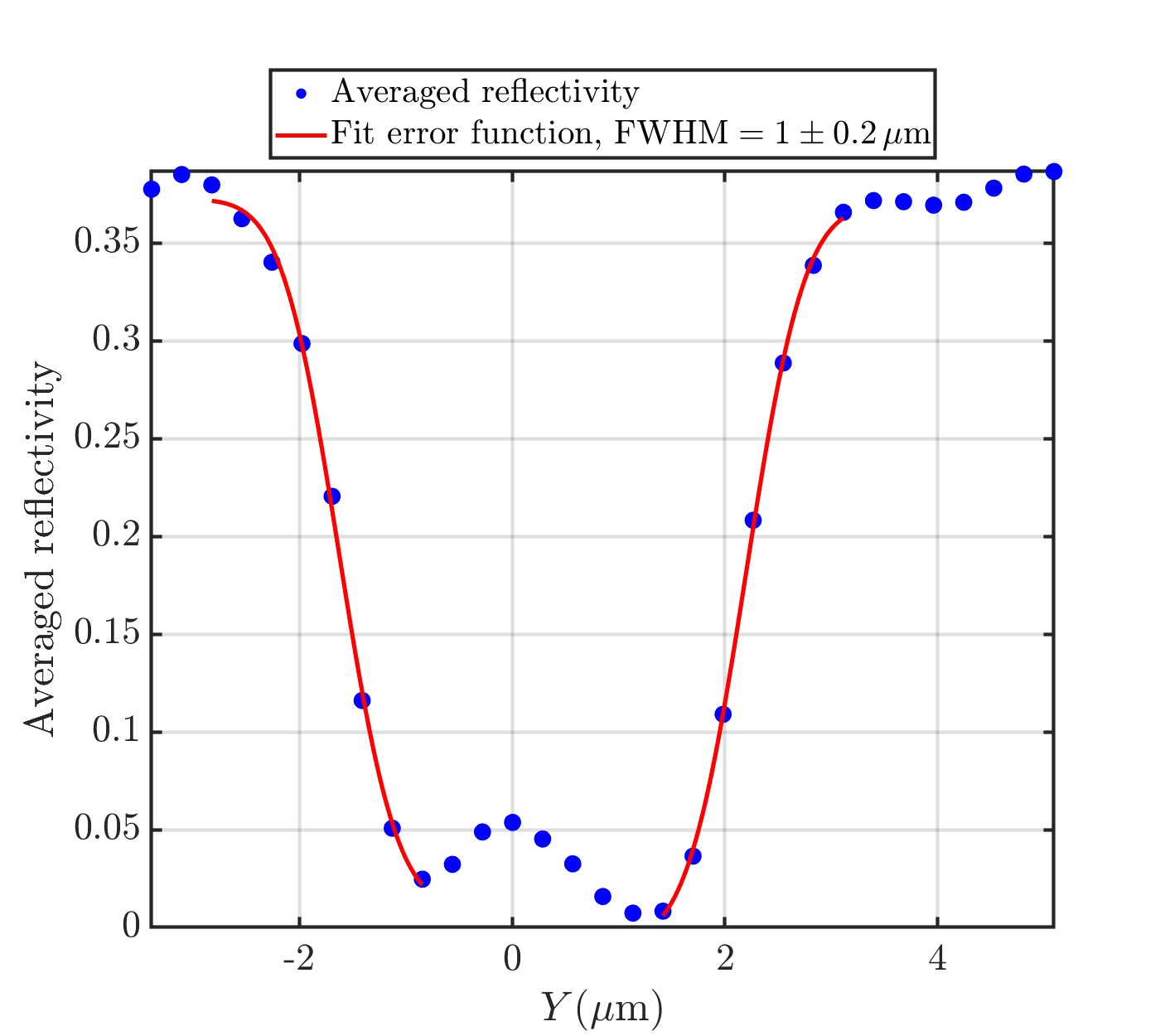}
         \caption{Averaged reflectivity of the data presented in Fig. \ref{YScan_DD} as a function of the $Y$-position and for $X = Z = 0$ in parallel polarization ($\hat{E}_{\mathrm{in}} = \hat{E}_{\mathrm{out}} = \hat{D}$).}
         \label{YScan_DD_Sum}
\end{figure}

To estimate the lateral resolution of the setup, we analyze the position scan of the cavity in parallel configuration. 
Fig. \ref{YScan_DD} in the main text depicts the reflection spectra as a function of $Y$ with $\hat{E}_{\mathrm{in}} = \hat{E}_{\mathrm{out}} = \hat{D}$. 
The reflectivity for each spectrum is averaged over the spectral range of the measurements ($0.87 - \SI{1.13}{\eV}$) and plotted as a function of $Y$ in Fig. \ref{YScan_DD_Sum}. 
The edges are fitted with an error function (a convolution of a Gaussian function and an edge function), which allows for determining the FWHM of the Gaussian resolution function.
From the fit in Fig. \ref{YScan_DD_Sum}, a FWHM of $\SI{1.0 \pm 0.2}{\micro\metre}$ is determined. 

For a diffraction-limited spot, the intensity in the focal plane is $\propto \left( 2 J_1(x)/x \right)^2 $, with $x = \frac{2 \pi}{\lambda} NA \, \rho$, where $NA$ denotes the numerical aperture ($= 0.65$ for our $50 \times$ objective) of the objective and $\rho$ denotes the lateral displacement \cite{Saleh1991}. This function has its first zero-value at $\rho = 0.61 \frac{\lambda}{NA}$, which is often referred to as the diffraction limit. The FWHM of this function is given by $\approx 0.51 \frac{\lambda}{NA} = \SI{0.87}{\micro\metre}$ at the photon energy of interest $E_{\mathrm{ph}} = \SI{1.116}{\eV}$ (the resonance energy of the high-$Q$ cavity mode). 

\newpage
\section{Far-field position scans}
\label{App_sec_posscan}

\begin{figure}[ht!]
         \centering
         \includegraphics[width = 0.6\linewidth]{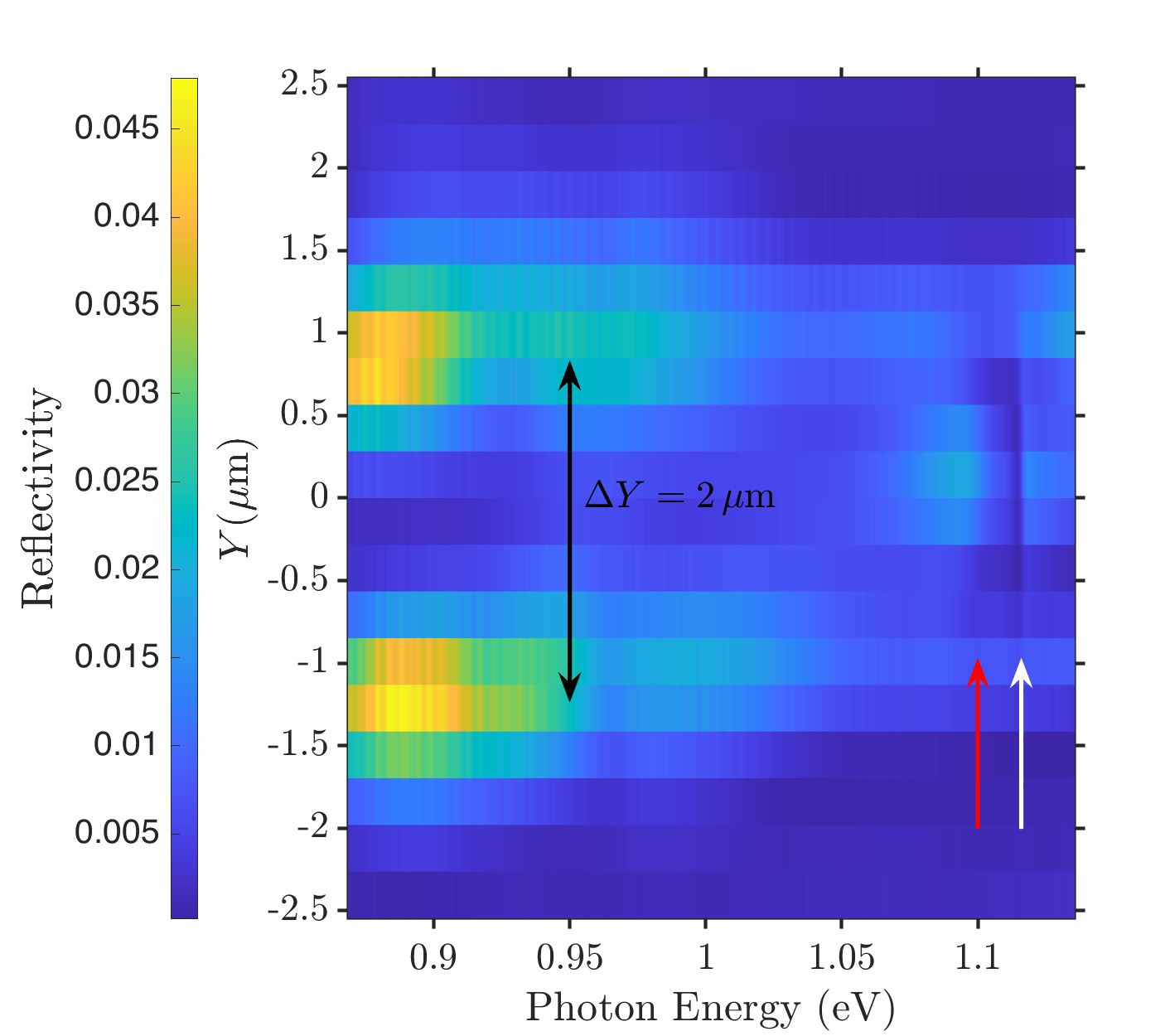}
         \caption{Reflection spectra as a function of the $Y$ position and for $X = Z = 0$ in the conventional cross-polarization configuration $\hat{E}_{\mathrm{in}} =\hat{A}$, $\hat{E}_{\mathrm{out}} = \hat{D}$. The high-$Q$ and low-$Q$ modes are marked with white and red arrows, respectively. The black arrow depicts the spatial distance of the maxima of the whispering gallery mode.}
         \label{Y_scan_crossPol}
\end{figure}

Fig. \ref{Y_scan_crossPol} depicts the full scan of the $Y$ position in cross-polarization ($\hat{E}_{\mathrm{in}} =\hat{A}$, $\hat{E}_{\mathrm{out}} = \hat{D}$) of the spectra discussed in Fig. \ref{Cross-pol spectra}. It can be seen that the high-$Q$ mode and low-$Q$ mode, marked with white and red arrows, respectively, are visible in the center of the cavity. The whispering-gallery mode is separated by $\approx \SI{2.0} {\micro\metre}$, marked by the black arrow, which fits well with the diameter of the outer rings of the cavity.

\begin{figure}[ht!]
    \centering
    \includegraphics[width=0.6\linewidth]{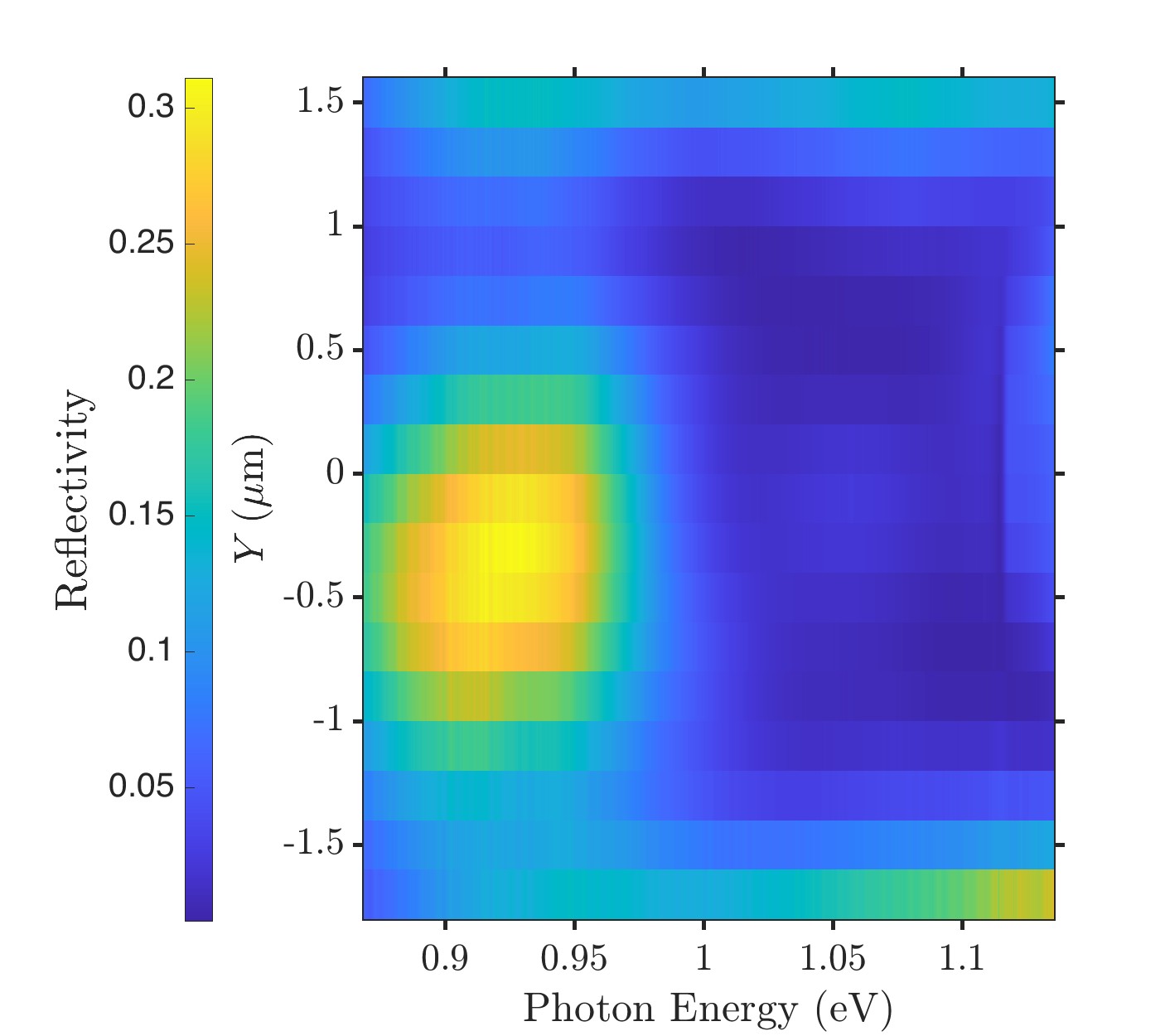}
    \caption{Reflection spectra of the cavity as a function of the $Y$-position and for $X = Z = 0$ in parallel polarization aligned with the high-$Q$ mode ($\hat{E}_{\mathrm{in}} = \hat{E}_{\mathrm{out}} = \hat{V}$) and without a $\lambda/4$ plate.}
    \label{YScan-VV}
\end{figure}

\begin{figure}[ht!]
    \centering
    \includegraphics[width=0.6\linewidth]{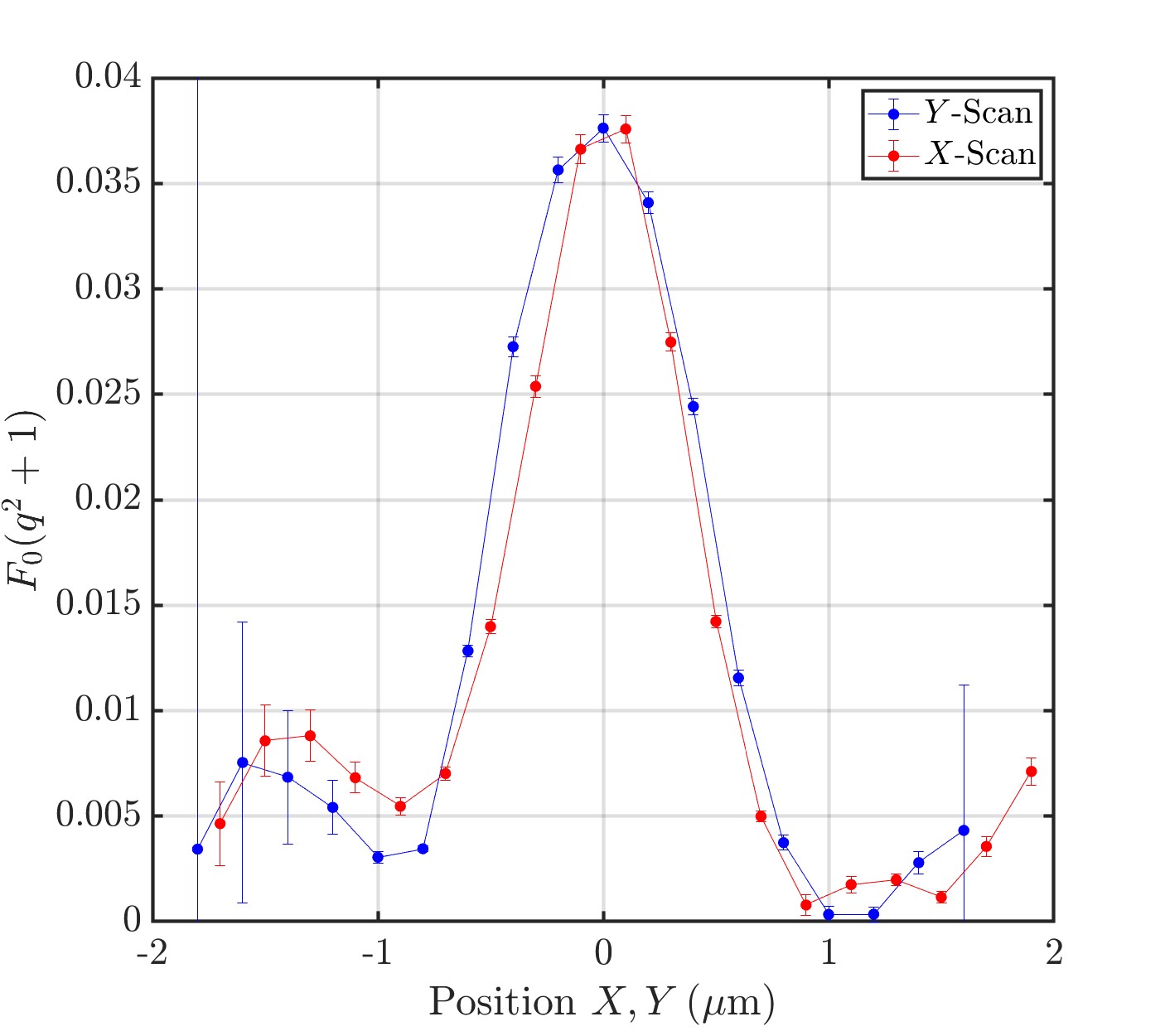}
    \caption{Extracted fit parameter $F_0(q^2+1)$ of the high-$Q$ mode for two spatial scans of the $X$ and the $Y$ positions, respectively (cf. Fig. \ref{YScan-VV}). Lines are guides to the eye.}
    \label{F0_VV}
\end{figure}

Fig. \ref{YScan-VV} depicts the reflected signal as a function of the $Y$-position with the input and output polarization aligned with the high-$Q$ mode (see main text). 
The high-$Q$ mode is clearly located in the center of the cavity. 
For further demonstration, we depict the extracted fit parameter for the amplitude $F_{0}(q^2+1)$ for two position scans along the $X$ and the $Y$ axis, cf. Fig. \ref{F0_VV}. 
For the various fits, the parameters $E_{0}$ and $\gamma$ were fixed to the values extracted from the fit in the center of the cavity, see Fig. \ref{FanoFit}. 
The mode amplitude is clearly centered in the cavity's center with a FWHM of $\approx \SI{1.0}{\micro\metre}$, which is exactly the extracted lateral resolution of the setup (see Sec. \ref{App_sec_latres}).

\begin{figure}[htb!]
    \centering
    \includegraphics[width=0.6\linewidth]{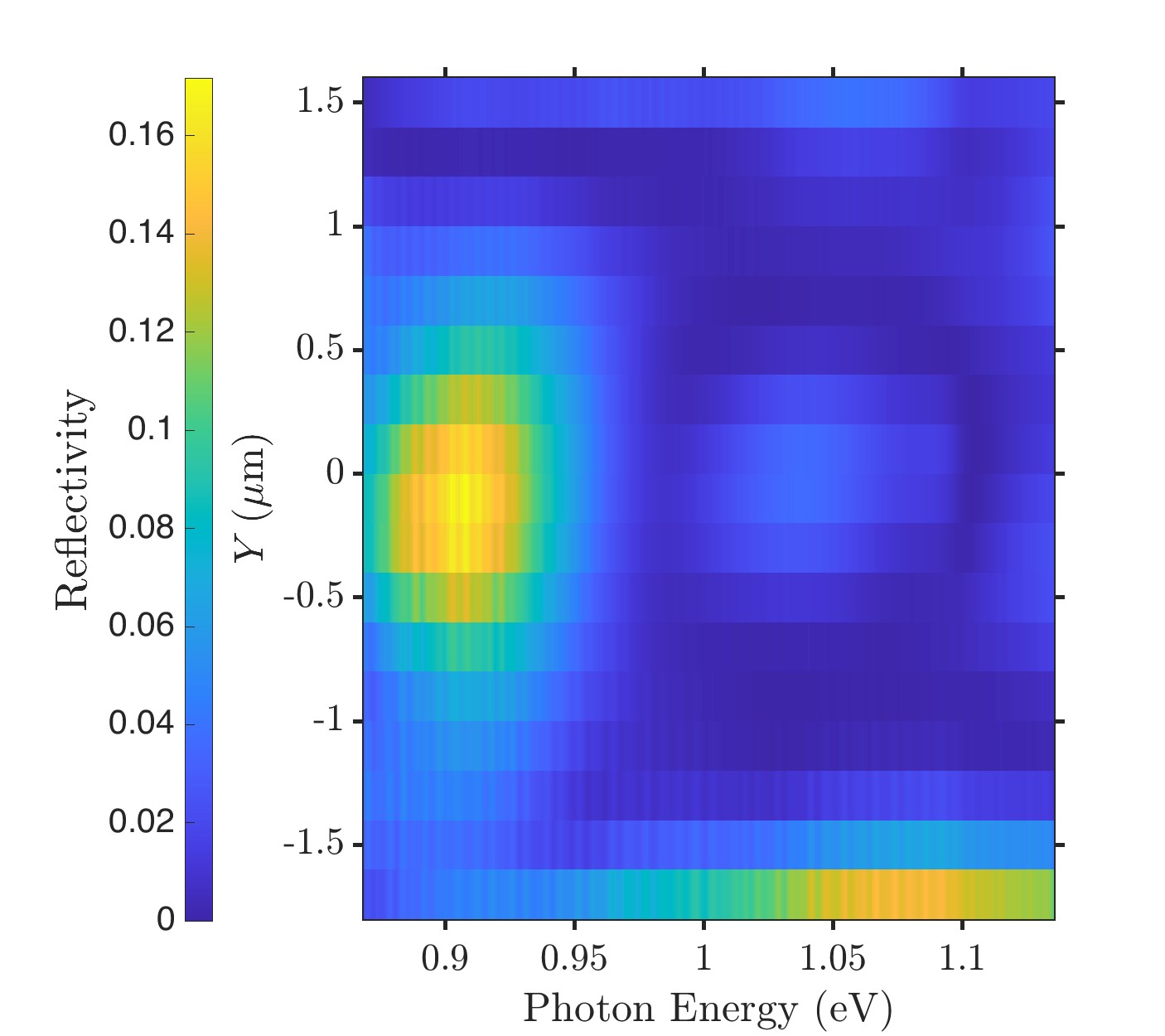}
    \caption{Reflection spectra of the cavity as a function of the $Y$-position and for $X = Z = 0$ in parallel polarization aligned with the low-$Q$ mode ($\hat{E}_{\mathrm{in}} = \hat{E}_{\mathrm{out}} = \hat{H}$).}
    \label{YScan_HH}
\end{figure}

\begin{figure}[ht!]
    \centering
    \includegraphics[width=0.6\linewidth]{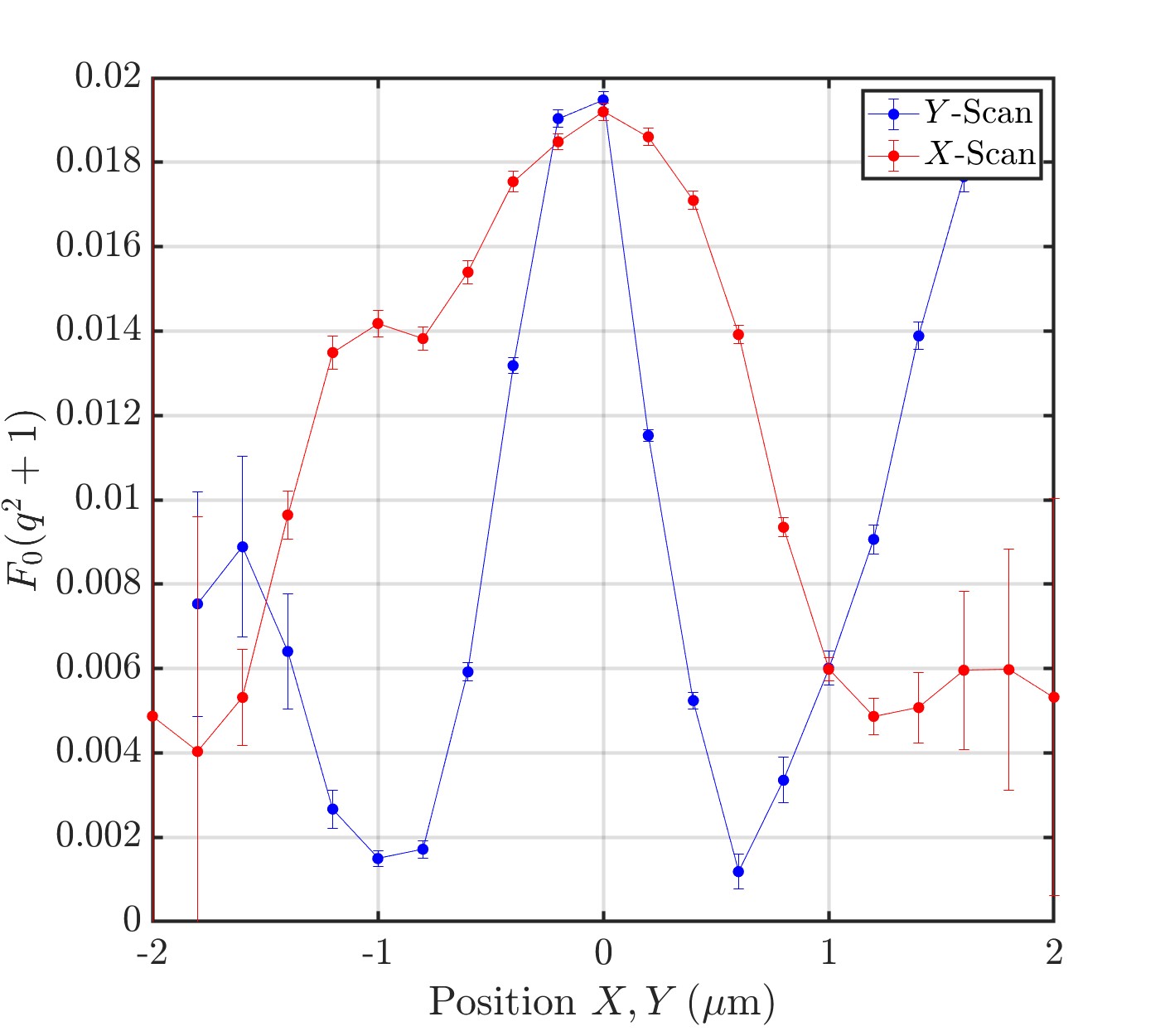}
    \caption{Extracted fit parameter $F_0(q^2+1)$ of the low-$Q$ mode for two spatial scans of the $X$ and the $Y$ positions, respectively (cf. Fig. \ref{YScan_HH}). Lines are guides to the eye.}
    \label{F0_HH}
\end{figure}

To investigate spatial properties of the low-$Q$ mode, position scans with $\hat{E}_{\mathrm{in}} =\hat{H}$, $\hat{E}_{\mathrm{out}} = \hat{H}$ are pursued. 
Fig. \ref{YScan_HH} depicts exemplary reflectivity as a function of the $Y$-position. 
The fit parameter $F_{0}(q^2+1)$ for the low-$Q$ mode as a function of the $X$ and the $Y$ positions is presented in Fig. \ref{F0_HH}. 
Here, the values for $E_{0}$ and for $\gamma_{0}$ were fixed as well. 
The low-$Q$ mode seems to be less confined than the high-$Q$ mode. 
In the $X$ direction, $F_{0}(q^2+1) (X)$ has a $\mathrm{FWHM}$ of $\approx \SI{1.8}{\micro\metre}$, compared to $\approx \SI{1.0}{\micro\metre}$ as for the high-$Q$ mode. 
In the $Y$ direction, the central part of the mode is sharply peak but 
$F_{0}(q^2+1) (Y)$ increases again for larger $Y$ values, which shows that the mode is even less confined in this direction. 
The reason for the asymmetry in the $Y$ direction is most likely a slight 
misalignment of the sample with respect to the symmetry axes of the cavity. 

\newpage
\
\newpage

\section{Near-field measurements}
\label{App_sec_sSNOM}

\begin{figure}[ht!]
    \centering
    \includegraphics[width=0.6\linewidth]{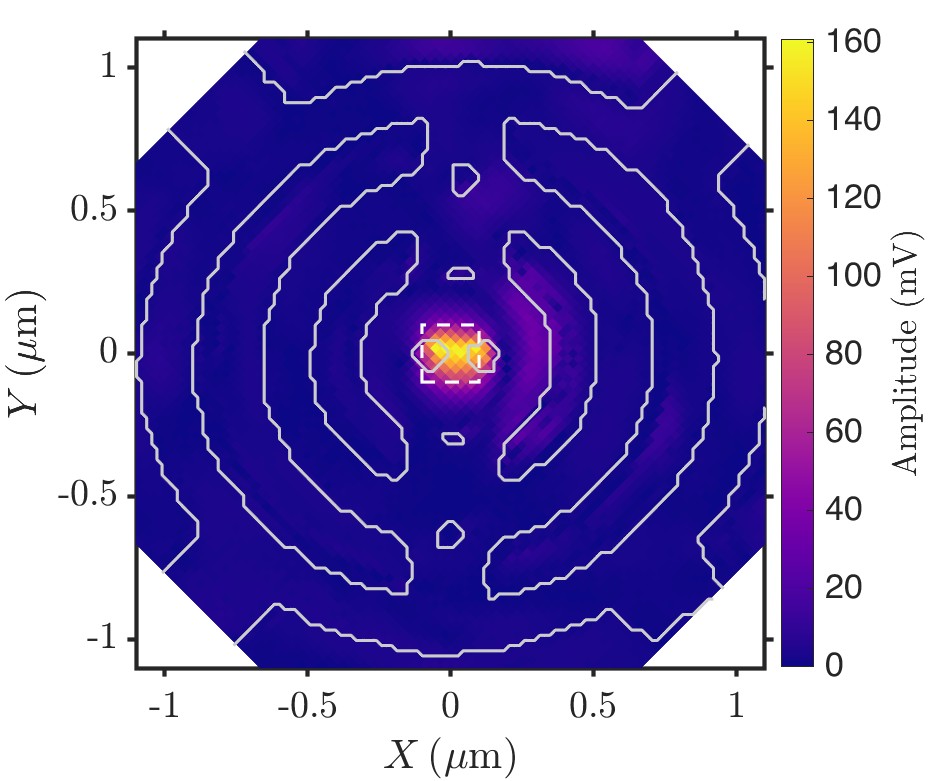}
    \caption{s-SNOM measurements of the 3rd-order scattering amplitude. The input and output polarizations are both along the $\hat{V}$ direction, aligned with the high-$Q$ mode \cite{Albrechtsen2022, Xiong2024}. The photon energy of a tunable laser is $\SI{1.105 }{\eV}$, aligned with the resonance of the high-$Q$ mode in the s-SNOM setup, cf. Fig. \ref{sSNOM_spec_highQ}. Solid contour lines depict the AFM profile of the EDC cavity. The dashed rectangle in the center depicts the spatial window for averaging of the amplitude to retrieve the spectrum.}
    \label{sSNOM_highQ}
\end{figure}

\begin{figure}[ht!]
    \centering
    \includegraphics[width=0.6\linewidth]{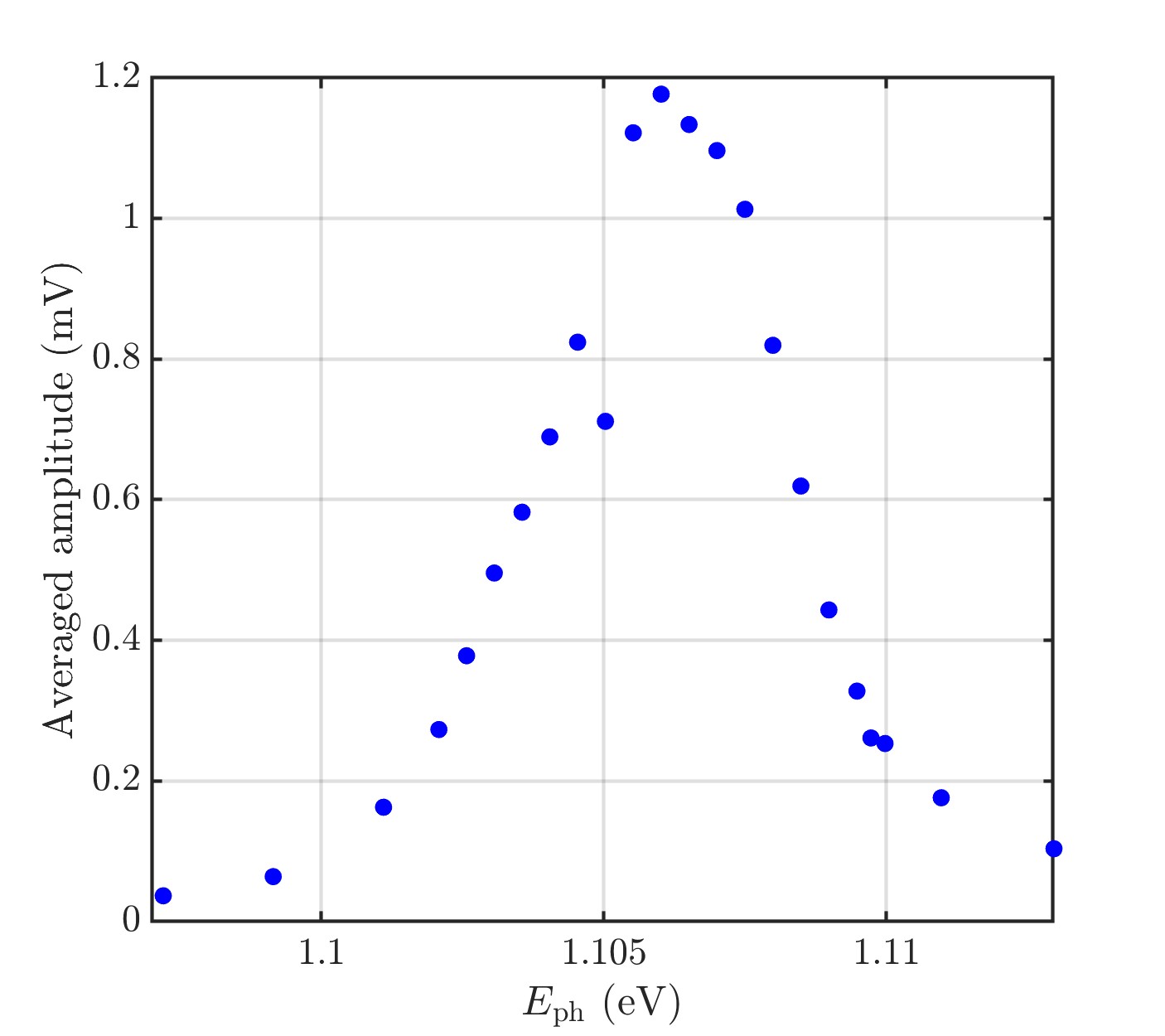}
    \caption{Average of 3rd-order scattering amplitude in a spatial window of $\SI{200}{\nano\metre}$ x $\SI{200}{\nano\metre}$, cf. Fig. \ref{sSNOM_highQ}. The input and output polarizations are both aligned with the $\hat{V}$ direction to detect the high-$Q$ mode. }
    \label{sSNOM_spec_highQ}
\end{figure}

Scattering-type scanning near-field optical microscope (s-SNOM) measurements \cite{Raschke2003} are often carried out to investigate the electric-field distribution in dielectric nanocavities \cite{Albrechtsen2022, Xiong2024}. In a s-SNOM, the apex of an atomic force microscope (AFM) tip, operating in tapping mode, is illuminated by a tunable laser \cite{Casses2024}. The detected scattered field is demodulated at higher harmonic orders $M$ of the tip-tapping frequency. Demodulation at $M \geq 3$ allows for retrieval of the near-field signal \cite{Knoll2000}. We present data demodulated at the 3rd harmonic order. Furthermore, pseudo-heterodyne detection \cite{Ocelic2006} is exploited to increase the near-field signal strength while suppressing the far-field signal.

We pursue s-SNOM measurements of the cavity with different polarizations and wavelengths. This way, we can retrieve the electric-field distribution of the high-$Q$ mode and gain valuable insight about the field distribution of the low-$Q$ mode. Fig. \ref{sSNOM_highQ} depicts the 3rd-order scattering amplitude of the cavity as a function of $X$ and $Y$. Here, the wavelength of a tunable laser is in resonance with the high-$Q$ cavity mode, see also Fig. \ref{sSNOM_spec_highQ}. As in previous studies \cite{Albrechtsen2022, Xiong2024}, the input and output polarizations are aligned in-plane and parallel to the high-$Q$ mode. Fig. \ref{sSNOM_spec_highQ} depicts the summed 3rd-order scattering amplitude in a spatial window of $\SI{200}{\nano\metre}$ x $\SI{200}{\nano\metre}$ in the center of the cavity. Clearly, a resonant behavior can be observed. The shift of $\approx \SI{10}{\milli\eV}$ of the resonance energy compared to far-field measurements (see Fig. \ref{FanoFit}) can be explained by the influence of the AFM tip on the cavity resonance.

Efficient near-field measurement of the low-$Q$ mode is not possible as the low-$Q$ mode is, to a large extent, located in a void region (see Fig. \ref{Sim_lowQ}). Thus, if the tip enters such a void region, the excitation laser light is shadowed by the surrounding dielectric material, so that solely scattering from the shaft of the AFM needle remains \cite{Albrechtsen2022, Xiong2024}.
\newpage
\
\newpage
\section{FEM simulations}
\label{App_sec_sim}

\begin{figure}[htb!]
    \centering
    \includegraphics[width=0.6\linewidth]{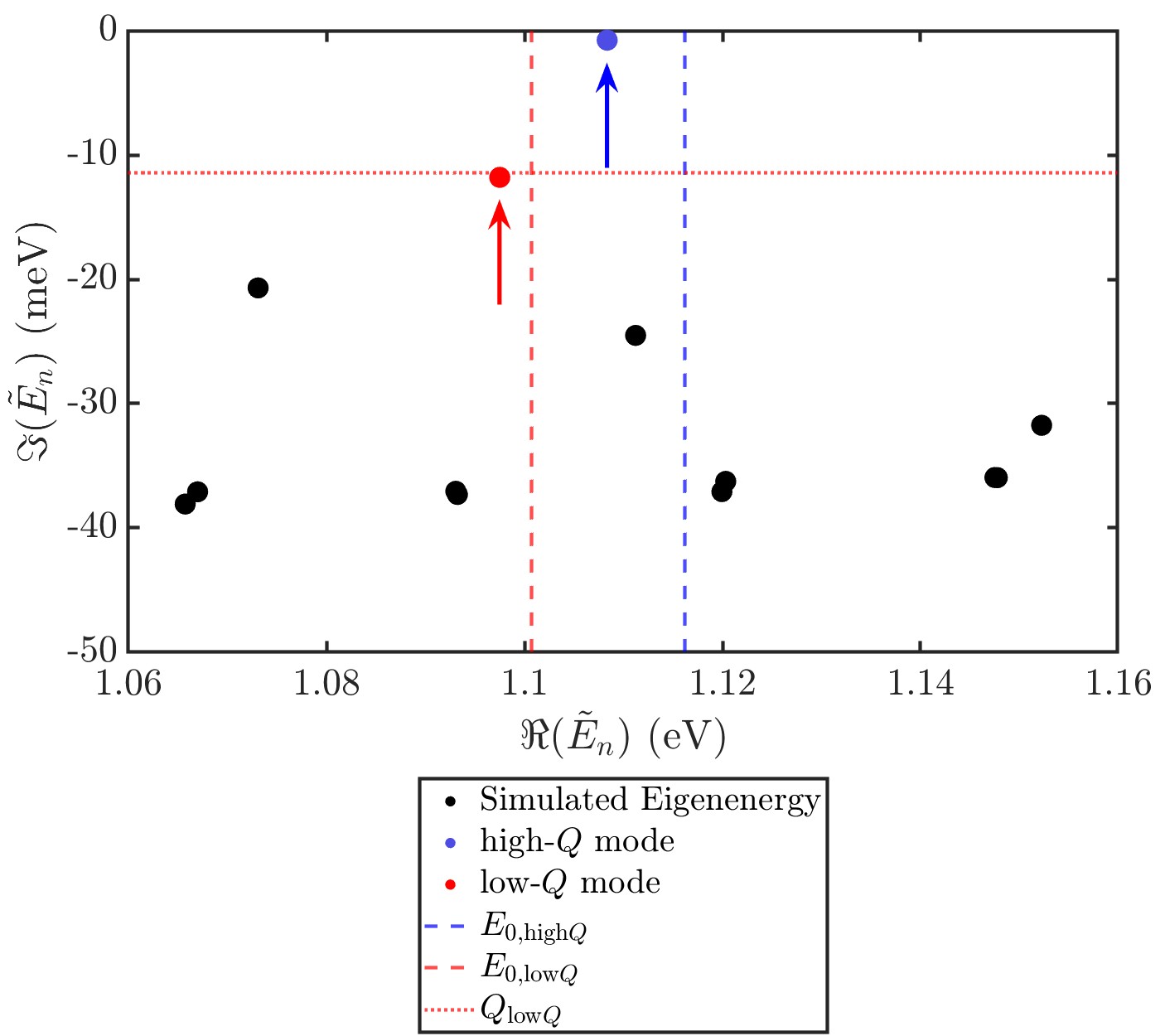}
    \caption{Complex eigenenergies $\tilde{E}_n$. The blue and red markers, together with the arrows, denote the simulated eigenenergy of the high-$Q$ and low-$Q$ modes, respectively. The blue- and the red-dashed lines mark the measured resonance energies of the high-$Q$ and of the low-$Q$ modes (cf. Sec. \ref{App_sec_fits}), respectively. The red-dotted line marks the decay rate corresponding to the $Q$-factor of the low-$Q$ mode.}
    \label{Complex_frequencies}
\end{figure}

The spectral and spatial information from the experiment are compared to finite-element-method simulation \cite{Kountouris2022}. We numerically solve the Helmholtz equation
\begin{equation}
    \nabla \times \nabla \times \tilde{\mathbf{f}}_n(\mathbf{r}) - \tilde{k}^2_{n}\epsilon_{\mathrm{R}}(\mathbf{r})\tilde{\mathbf{f}}_n(\mathbf{r}) = 0
\end{equation}
with scattering boundary conditions. $\tilde{\mathbf{f}}_n(\mathbf{r})$ denotes the electric field of the eigenmode, $\tilde{k}^2_{n} = \tilde\omega_n/c$ the corresponding wavenumber with the eigenfrequency $\tilde\omega_n$ and $\epsilon_{\mathrm{R}}(\mathbf{r})$ the relative permittivity. This way, we retrieve the eigenmodes of the system in the quasinormal mode framework \cite{TrøstKristensen2021}. We restrict our calculation to the InP membrane surrounded by air, and apply sufficient symmetry conditions for the high-$Q$ and the low-$Q$ mode, respectively (see below).

Fig. \ref{Complex_frequencies} shows the simulated eigenenergies $\tilde{E}_n = \hbar \tilde{\omega}_n$. The values for the high-$Q$ mode ($\tilde{E}_{\mathrm{high}Q}$) and the low-$Q$ mode ($\tilde{E}_{\mathrm{low}Q}$) have been calculated with a convergence study as described in Ref. \cite{Kountouris2022}. Those values are indicated with blue and red markers and arrows, respectively. The calculated real part of the eigenenergies is red-detuned from the value observed experimentally, indicated by the blue and red dashed lines, for both modes. This can be explained by fabrication imperfections. As described in Ref. \cite{Kountouris2022}, the resonance energy is very sensitive to slight variations in the geometry. We find that decreasing the radius of the holes in the center of the cavity by $\SI{3}{\nano\metre}$ decreases the real part of the resonance energy of the high-$Q$ mode by approximately $\SI{20}{\milli\eV}$. That is why we adjust the nominal geometry according to the SEM image (Fig. \ref{Sample}). As the cavity depicted in Fig. \ref{Sample} is a clone of the studied cavity, small variations of the geometry might still appear. Moreover, surface roughnesses and oxidation effects can influence the resonance energies. 

\begin{figure}[htb!]
    \centering
    \includegraphics[width=0.6\linewidth]{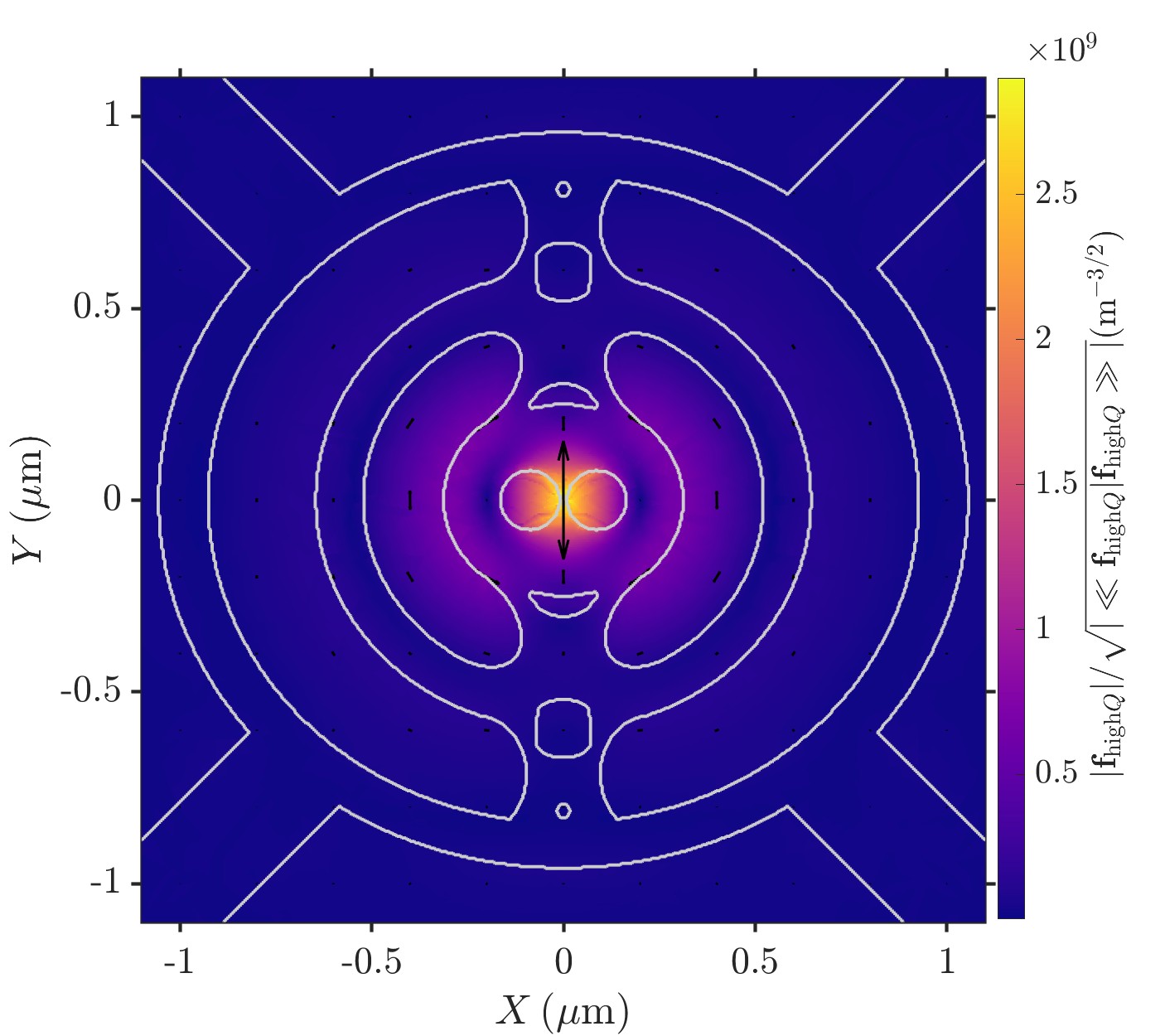}
    \caption{Normalized field profile of the high-$Q$ mode, evaluated in a plane $\SI{5}{\nano\metre}$ above the EDC cavity. The white line draws the contour of the EDC cavity. The black arrows denote the in-plane orientation of the electric field.}
    \label{Sim_highQ}
\end{figure}

We find $\mathfrak{R}(\tilde{E}_{\mathrm{high}Q}) = (1.10835\pm2\times10^{-4})$ eV, which is around $\SI{8}{\milli\eV}$ smaller than the experimental value $E_{0,\mathrm{high}Q} = \SI{1.1162\pm0.0001}{\eV}$, cf. Fig. \ref{FanoFit}. The calculated $Q$-factor of the high-$Q$ mode is $723 \pm 6$, which is much larger than the measured $Q_{\mathrm{high}Q} = 265 \pm 8$, a discrepancy which is very common in dielectric nanocavities \cite{Albrechtsen2022, Xiong2024}. The simulated resonance energy of the low-$Q$ mode $\Re(\tilde{E}_{\mathrm{low}Q}) = \SI{1.0975\pm0.0002}{\eV}$ differs by $\SI{3}{ \milli\eV}$ from the measured value $E_{0,\mathrm{low}Q} = \SI{1.1007\pm0.0003}{\eV}$. Moreover, the calculated $Q$ factor for the low-$Q$ mode is $46.6\pm0.4$. This value agrees well with the measured value $Q_{\mathrm{low}Q} = 48 \pm 1$, indicated by the dotted line in Fig. \ref{Complex_frequencies}.

The normalized field profile $|\mathbf{f}_{\mathrm{high}Q}|/\sqrt{|\langle\langle \mathbf{f}_{\mathrm{high}Q}|\mathbf{f}_{\mathrm{high}Q} \rangle\rangle|}$ evaluated in a plane $\SI{25}{\nano\metre}$ above the EDC cavity corresponding, to the effective scattering distance between of the s-SNOM tip and the sample surface \cite{Xiong2024}, is shown in Fig. \ref{Sim_highQ}. The mode is located in the center of the EDC cavity, which is verified with polarization tomography (Fig. \ref{F0_VV}) and near-field measurements, see Sec. \ref{App_sec_sSNOM}. The in-plane orientation of the electric field demonstrates polarization mainly along the $\hat{V}$ direction, matching experimental observations, see Sec. \ref{sec:results}. The high-$Q$ mode is symmetric with a perfect electric conductor (PEC) boundary condition on the $XZ$-plane and a perfect magnetic conductor (PMC) boundary condition on the $YZ$-plane, as well as a PMC layer on the $XY$-plane \cite{Kountouris2022}. 

\begin{figure}[htb!]
    \centering
    \includegraphics[width=0.6\linewidth]{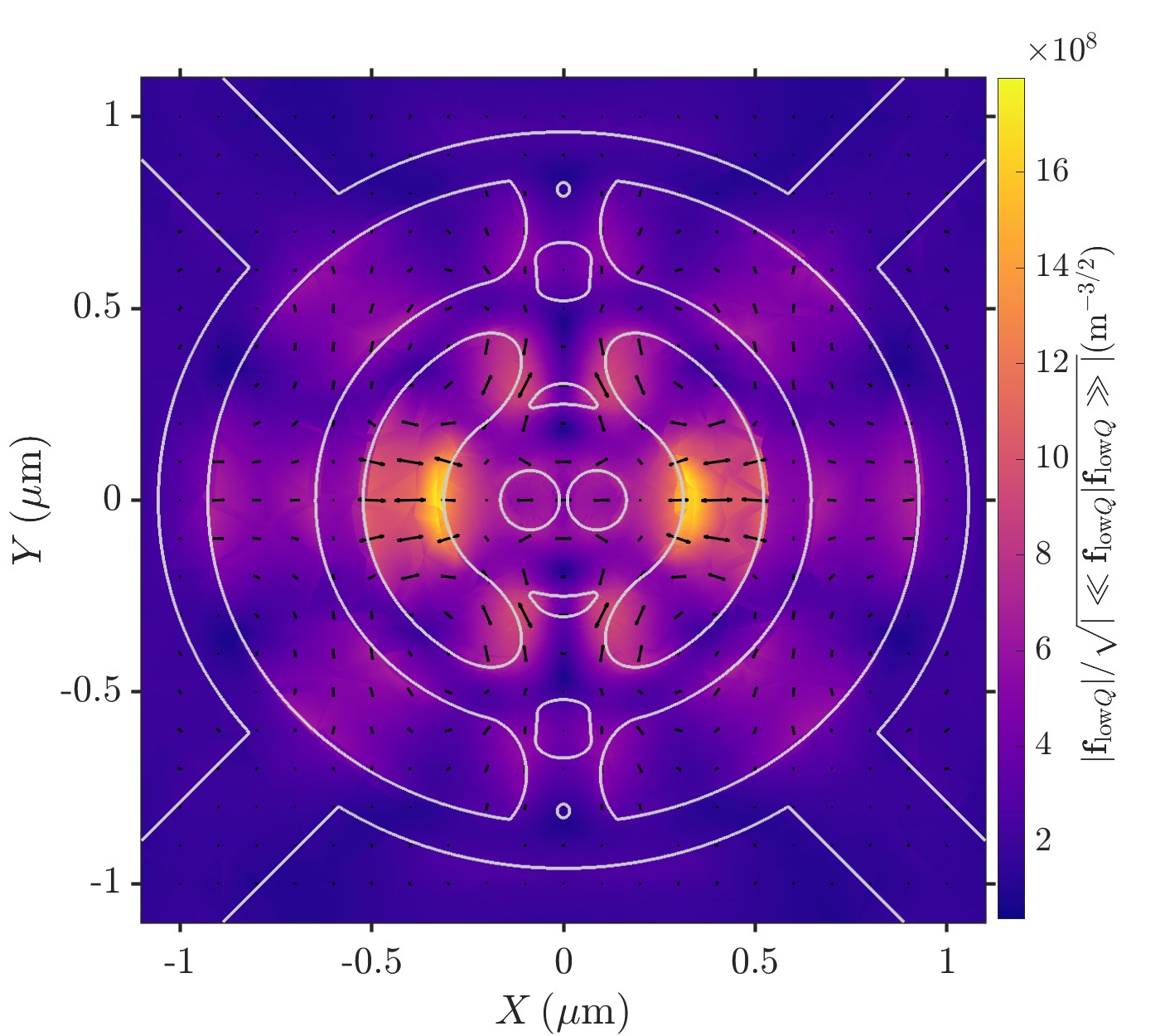}
    \caption{Normalized field profile of the low-$Q$ mode, evaluated in a plane $\SI{5}{\nano\metre}$ above the EDC cavity. The white line draws the contour of the EDC cavity. The black arrows denote the in-plane orientation of the electric field.}
    \label{Sim_lowQ}
\end{figure}

The normalized field profile $|\mathbf{f}_{\mathrm{low}Q}|/\sqrt{|\langle\langle \mathbf{f}_{\mathrm{low}Q}|\mathbf{f}_{\mathrm{low}Q} \rangle\rangle|}$ evaluated in a plane $\SI{25}{\nano\metre}$ above the EDC cavity is shown in Fig. \ref{Sim_lowQ}. The mode is extended in the $X$- and $Y$-directions, matching polarization tomography measurements, see Fig. \ref{F0_HH}. From the in-plane orientation of the electric field, it follows that the mode is mainly polarized along the $\hat{H}$ direction, matching experimental observations, see Sec. \ref{sec:results}. The low-$Q$ mode is symmetric with a PMC layer on the $XZ$-plane and a PEC layer on the $YZ$-plane, as well as a PMC layer on the $XY$-plane.

\newpage
\bibliographystyle{unsrt}
\bibliography{PRR}

\end{document}